\documentclass[pdflatex,sn-nature]{sn-jnl}

\usepackage{graphicx}%
\usepackage{multirow}%
\usepackage{amsmath,amssymb,amsfonts}%
\usepackage{amsthm}%
\usepackage{mathrsfs}%
\usepackage[title]{appendix}%
\usepackage{xcolor}%
\usepackage{textcomp}%
\usepackage{manyfoot}%
\usepackage{booktabs}%
\usepackage{algorithm}%
\usepackage{algorithmicx}%
\usepackage{algpseudocode}%
\usepackage{listings}%
\usepackage{xr}
\theoremstyle{thmstyleone}%
\theoremstyle{thmstyletwo}%

\theoremstyle{thmstylethree}%

\begin{document}

\title[Article Title]{Anatomy of a Complex Crystallization Pathway}


\author[1]{\fnm{Charlotte Shiqi} \sur{Zhao}}\email{zshiqi@umich.edu}

\author[1]{\fnm{Domagoj} \sur{Fijan}}\email{dfijan@umich.edu}

\author*[1,2]{\fnm{Sharon} \sur{C. Glotzer}}\email{sglotzer@umich.edu}

\affil[1]{\orgdiv{Department of Chemical Engineering}, \orgname{University of Michigan}, \orgaddress{\street{2800 Plymouth Road, Building 10}, \city{Ann Arbor}, \postcode{48109}, \state{Michigan}, \country{United States}}}

\affil[2]{\orgdiv{Biointerfaces Institute}, \orgname{University of Michigan}, \orgaddress{\street{2800 Plymouth Road, Building 10, A175}, \city{Ann Arbor}, \postcode{48109}, \state{Michigan}, \country{United States}}}


\abstract{Using molecular dynamics simulations, we investigate the crystallization pathways of two exemplary systems that form the same complex crystal structure but differ fundamentally in the nature of their particle interactions. One system is composed of point particles interacting via an isotropic pair potential characteristic of metallic compounds, while the other system contains hard polyhedra whose interactions arise from emergent entropic forces. Despite the stark difference in the origins of the particle interactions, we find that both systems are polymorphic and share the same crystal polymorphs. Moreover, the two systems follow the same multistep crystallization pathways, and by examining the complex crystallization pathways on the single particle level, we find that the local structure evolution of the two systems is also similar. By mapping the hard particle system’s interaction to an effective pairwise potential, we find that such resemblance arises from the particle interactions being effectively similar.
}




\maketitle

\section{Introduction}\label{intro}

Rational design of nanoparticles is a key step towards the realization of novel nanomaterials~\cite{boles2016self, lee2022nanoparticle}. Compared to atoms, whose behaviors are dictated by electronic structure, nanoparticles provide virtually unlimited tunability in terms of valency and their behavior can be prescribed through intentional engineering of their attributes~\cite{glotzer2007anisotropy}. Therefore, much effort has been devoted to synthesizing nanoparticle building blocks and an impressive level of control has been achieved in both shape and interaction anisotropy~\cite{chen2019recent, pearce2021synthesis,hueckel2021total, nguyen2022colloidal, chen2024nanoparticle, ibanez2025prospects}. Such patchy particles can now be self-assembled into structures of remarkable complexity~\cite{wang2023controlled, zhou2024colloidal, zhou2024engineering}.

Along with progress in particle design and synthesis, advances are also being made to control the self-assembly pathways of nanoparticles such that structures with desired functionality can be selectively fabricated~\cite{luo2023unravelling, ou2020kinetic,kim2017imaging,gentili2022reversible,paulson2015control,akkerman2022controlling, zhou2022chiral, zhong2024engineering}. This control is especially important for complex colloidal crystals with large unit cells~\cite{lee2023entropy,lu2025octo, ye2015structural,lee2014sphericity,lin2017clathrate, reddy2018stable, bommineni2019complex, su2019identification} and crystals with competing polymorphs~\cite{nozawa2025polymorphic}. While our understanding of self-assembly pathways is increasing, much remains to be explored and there are many open questions~\cite{boles2016self, du2024non, bassani2024nanocrystal, jacobs2025assembly}. Experimentally, it is still challenging to directly observe and examine assembly pathways despite recent advances in liquid-phase transmission electron microscopy~\cite{ou2020kinetic, luo2023unravelling, kim2024recent, zhong2024engineering}. Here, simulations can provide important insight by probing the dynamics with single-particle level resolution.

Using simulation, we seek to study and compare the crystallization pathways of complex-crystal-forming systems with disparate particle interactions. Complex-crystal-forming systems of particles are particularly interesting because many of the structures self-assembled by nanoparticles can be found in systems across many length scales where the interactions of the components are fundamentally different~\cite{bassani2024nanocrystal, dshemuchadse2022soft, zhou2024colloidal}. For example, soft spherical micelles formed by block copolymers~\cite{lee2010discovery} can crystallize into the $\sigma$-phase, an example of a complex topologically close-packed structure commonly adopted by metal alloys~\cite{frank1958complex, frank1959complex}. Additionally, it has been shown in computer simulations that hard tetrahedra and hard pentagonal orthobicupola, which have a disk-like shape, can also self-assemble into the $\sigma$-phase via emergent entropic forces~\cite{haji2011phase, damasceno2012predictive, van2014understanding}. Elucidating the formation mechanisms for identical crystal structures formed by different classes of building blocks may shed light on the crystallization process of their counterparts in other disparate systems. 

Another example is the clathrate structure. Commonly occurring as gas hydrates, clathrates are composed of polyhedral cages that can host guest components. A variety of guest-free~\cite{lin2017clathrate} and host-guest~\cite{lee2023entropy} colloidal clathrate structures have been achieved in experiments and simulations of DNA-modified and hard, truncated triangular bipyramidal nanoparticles, respectively. Remarkably, the crystallization pathway of one hard particle clathrate, which assembles solely due to entropy maximization, shares key features in common with clathrate formation of water molecules~\cite{lee2019entropic}. 

Inspired by these findings, we seek to answer the following types of questions: What is the origin of the similarity in crystallization pathways for systems of particles that form the same crystal structures but with very different types of interactions? Are the pathways the same simply because the crystal structures are the same? If so, do other disparate systems that form the same complex crystal follow similar pathways as well?

To investigate these questions, we performed molecular dynamics simulations to study two exemplary model systems that form the same complex crystal structure, and compare their crystallization pathways. The two systems differ in the nature of their interactions. The first system is composed of point particles interacting via an isotropic pair potential designed by Zetterling et al.~\cite{zetterling2000gamma}. The multiwells of this potential are characteristic of interactions of metal atoms in intermetallic compounds. The second system consists of hard truncated tetrahedra (TT) whose interactions are anisotropic and solely entropic~\cite{klotsa2018intermediate}.
Both systems have been reported to crystallize into the same complex crystal structure, $\gamma$-Brass (\textit{cI52-$\text{Cu}_{5}\text{Zn}_{8}$})~\cite{zetterling2000gamma, klotsa2018intermediate}. $\gamma$-Brass phases are a class of metallic alloys, and the cubic $\gamma$-Brass structure has 52 atoms in its unit cell that are divided among four Wyckoff sites~\cite{lord2004gamma, gourdon2007atomic}. $\gamma$-Brass nanocrystals have been reported to show great promise as catalysts for various processes~\cite{dasgupta2019generalized}. 

Here, our interest in the crystallization pathways of $\gamma$-Brass is motivated not by its potential for practical applications, but by the fact that such complex structural order emerges from two disparate systems, making this particular structure a good candidate for our purposes. Through extensive molecular dynamics simulations, we report that each system can crystallize into four competing polymorphs at the thermodynamic state point at which $\gamma$-Brass forms, three of which are common to both systems. We identified the multistep transitions among the polymorphs exhibited by both systems. To examine these transitions closely, we used a simple machine learning method to dissect the complex crystallization pathways down to the single-particle level, which reveals that the emergence and subsequent evolution of local structures (LSs) in both systems also exhibit similarities between the two systems. By comparing the structure of the pre-nucleation fluids and analyzing the associated potential of mean force, we demonstrate that the interactions are effectively similar at the thermodynamic state points of interest, thus giving rise to the isopolymorphism and similar pathways, even though in one system crystallization is driven by potential energy minimization and the other by entropy maximization. We posit that this parallel is not unique to these two systems, and instead indicates a more universal connection between purely entropic systems and systems with enthalpic, or both entropic and enthalpic, particle interactions.

\section{Results}\label{results}

\subsection{Similar Polymorphism in Disparate Systems} \label{subsec:poly}

\begin{figure*}[t]
    \includegraphics[width=1\linewidth]{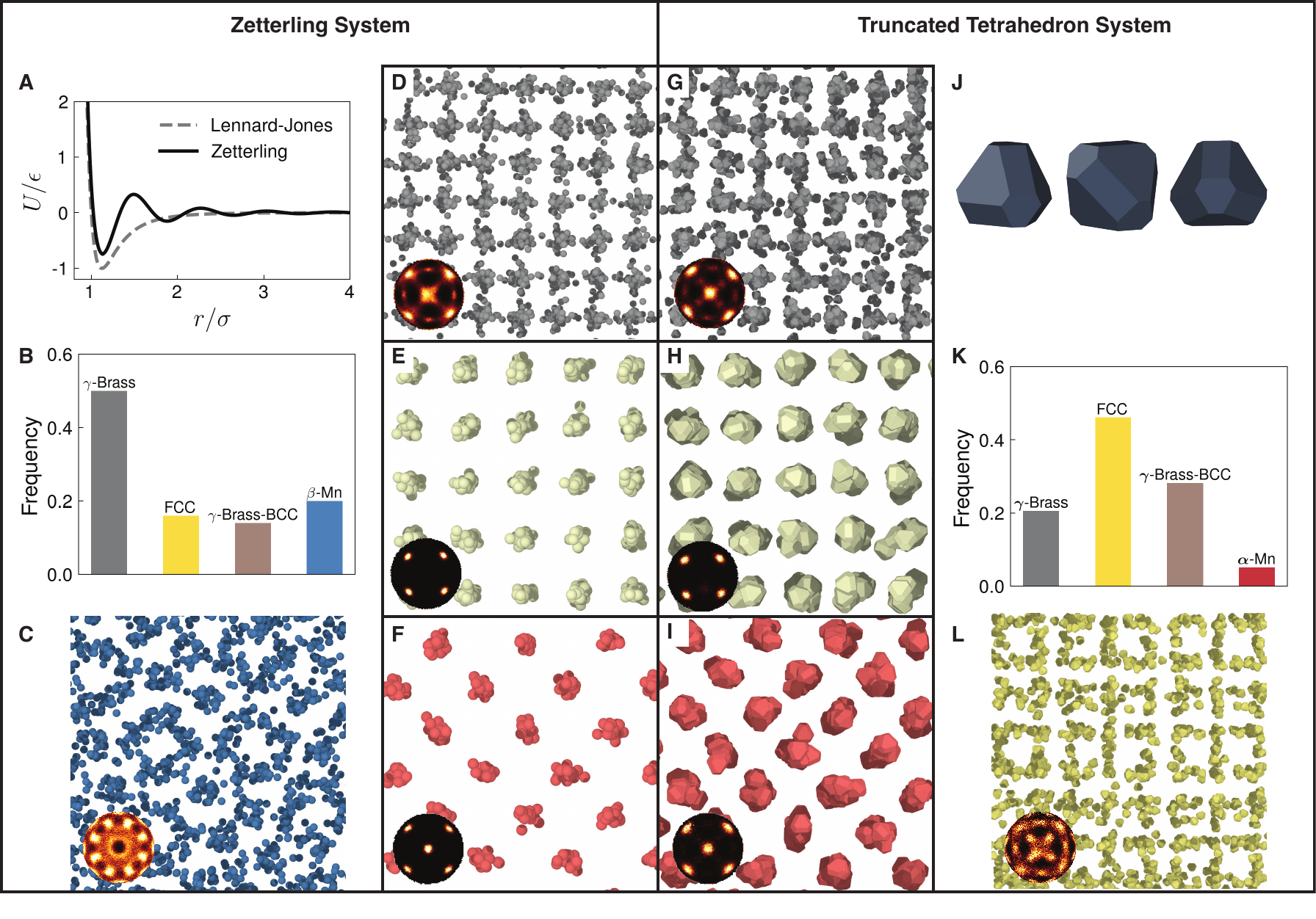}
    \caption{(A) The Zetterling pair potential (the Lennard-Jones potential is shown for comparison) and (J) the truncated tetrahedron shown in different views. Frequency of the crystal polymorphs obtained in our simulations for the (B) Zetterling system and the (K) TT system. Crystal polymorphs formed by the Zetterling system (C-F) and the TT system (G-I, L) shown with the corresponding BODs: (C) $\beta$-Mn, (D, G) $\gamma$-Brass, (E, H) FCC, (F, I) BCC, and (L) $\alpha$-Mn. Note that the TT shapes in G, H, I and L are displayed at a smaller scale to showcase the crystal lattices.
   }
    \label{fig:polymorphism}
\end{figure*}

We performed extensive molecular dynamics simulations for both the Zetterling potential and hard particle systems at a thermodynamic state point at which the systems form $\gamma$-Brass from the fluid phase spontaneously (see the Methods section for details). The Zetterling potential and the truncated tetrahedral shape we used are shown in Fig.~\ref{fig:polymorphism} A and J, respectively.

For the Zetterling system, we focus on the results of $50$ independent simulations of $N = 80, 000$ particles each. In these simulations, we found that the system can form any one of the following structures (Fig.~\ref{fig:polymorphism}): mixtures of BCC and $\gamma$-Brass, FCC with HCP defects, $\beta$-Mn (which has $20$ particles in its unit cell distributed among two Wyckoff sites), and $\gamma$-Brass -- the most frequently observed polymorph. Interestingly, although the Zetterling system is capable of forming large BCC crystal grains, they only appear in mixture phases of BCC and $\gamma$-Brass (Fig.~\ref{fig:polymorphism}B), and are usually found between layers of $\gamma$-Brass or surrounded by $\gamma$-Brass grains. 
We tested the stability of BCC with various system sizes ranging from $N = 6,750$ to $N = 85750$, and the system remains stable in BCC in all our stability tests (Table~\ref{tab:zetterling_stability_result}); we found that it has an even lower Gibbs free energy than $\gamma$-Brass (Fig.~\ref{fig:stability}A). However, only in the assembly simulations with fewer particles ($8,000$) did the entire system form a BCC crystal; it may be that the formation of $\gamma$-Brass is inhibited by finite size effects. The metastability of the mixture phase of BCC and $\gamma$-Brass likely stems from the close structural relationship between the two crystals. More specifically, $\gamma$-Brass can be viewed as a modified BCC lattice~\cite{lord2004gamma, gourdon2007atomic}.

Our free energy calculations show that FCC has the lowest Gibbs free energy among the polymorphs formed by the Zetterling system (Fig.~\ref{fig:stability}A), but it is not the most frequently observed crystal form. In only $8$ out of the $50$ simulations did the entire system form an FCC crystal (Fig.~\ref{fig:polymorphism}B), which indicates high energy barriers between the fluid phase and FCC.

For the TT system, we performed $40$ independent molecular dynamics simulations with $N = 10,000$ particles each, and here we focus on the results of $39$ of them, as one of the 40 systems failed to self-assemble into any crystal. In the $39$ simulations, we found that the system can form the following phases (Fig.~\ref{fig:polymorphism}): $\gamma$-Brass, FCC with HCP defects, mixtures of $\gamma$-Brass and BCC, and $\alpha$-Mn (whose unit cell has $58$ particles divided among four Wyckoff sites).

Similar to the Zetterling system, FCC also has a lower free energy than $\gamma$-Brass in the TT system (Fig.~\ref{fig:1D_freeE}). However, unlike Zetterling, FCC is the most frequently observed crystal polymorph for the TT system (Fig.~\ref{fig:polymorphism}K). As will be discussed below, in most cases, the system still forms $\gamma$-Brass first, regardless of the final equilibrium crystal structure. This observation is supported by hybridMC-MetaD ~\cite{zhao2026hybrid} simulation results for the TT system: from the $\gamma$-Brass phase, the system needs to overcome an energy barrier of only around $10k_BT$ to form FCC (Fig.~\ref{fig:1D_freeE}), which is even lower than the barrier separating fluid and $\gamma$-Brass (around $14k_BT$). In other words, the formation of $\gamma$-Brass facilitates the formation of FCC in the TT system.

\subsection{Local Structure Classification} \label{subsec:classify}

To investigate the transition pathways among the polymorphs on the single-particle level, we quantify the local structures (LSs) in the systems by computing a set of local bond orientational order parameters and the Voronoi density for each particle (see the \hyperref[method:features]{Local Structure Features} section for more details) and track the evolution of the LSs.

Machine learning (ML) methods, both supervised and unsupervised, have proved to be especially useful for such studies~\cite{boattini2019unsupervised, adorf2019analysis, boattini2020autonomously, coli2021artificial, de2023search, martirossyan2024local}. Given the rapid advances in artificial intelligence and the growing body of research leveraging ML methods in this field, we detail the rationale behind our methodological choice here, hoping to offer guidance for future work. 

For pathways that involve many types of LSs, which is the case for the two systems studied in this work, unsupervised clustering methods tend to generate results that can be hard to interpret. For example, it can be difficult to associate the clusters with relevant local structures~\cite{adorf2019analysis}. Another common issue in studying LSs in complex crystals is that these crystals have polytetrahedral and icosahedral local order that is abundant in the supercooled liquids from which the crystals form~\cite{nelson1989polytetrahedral}. In other words, several of the phases involved have intrinsically similar LSs, and we found that an unsupervised clustering approach struggles to distinguish the LSs of these phases. The presence of defects in the assembled crystals further complicates the issue.

One strong suit of unsupervised ML methods is the minimal assumptions about the system that are needed and thus unsupervised ML has been used to search for hidden structural order~\cite{de2023search}. However, through our preliminary investigation of both systems, we found no evidence of precursors that are structurally different from the observed crystal polymorphs. Furthermore, unsupervised clustering methods still require an estimation of the number of clusters, which again introduces subjectivity. For these reasons, we instead used a supervised ML method with the following assumption: the LSs of the two systems studied here can be fully described by the LSs in the relevant crystals and the LSs in the fluid phases. Although some of the LSs in the complex crystals and the fluid are similar, we include an extra "fluid" type to account for the LSs that deviate greatly from the complex-crystal LSs. We also assume that the LSs in $\beta$-Mn and $\alpha$-Mn can be found in both systems, even though neither crystal is formed by both systems. We thus categorize the LSs for both systems into the following seven types: fluid, BCC, FCC, HCP (appeared as defects in the self-assembled FCC crystals), $\gamma$-Brass, $\beta$-Mn and $\alpha$-Mn.

Due to their intrinsic similarities, the LSs in the fluid phase and the complex crystals are also in close proximity in the feature space and are separated by highly nonlinear boundaries. The $k$-Nearest Neighbors ($k$NN) algorithm~\cite{cover1967nearest} is thus an intuitive choice for our classification task, as it is a distance-based nonlinear classification method, in which unseen samples are assigned the same class as the most frequent one among their $k$ nearest neighbors in feature space. We trained two $k$NN classifiers, with which we first label the particles as crystal- or fluid-like in the self-assembly trajectories, and then further identify the LS types for the crystal-like particles. For the $k$NN algorithm, we only need to choose one hyperparameter, $k$, and we found that the performance of the $k$NN classifiers is fairly robust against different $k$ values (Table~\ref{tab:zetterling_score} and Table~\ref{tab:tt_score}), owing to our choice of features and carefully curated training dataset (see \textit{SI Appendix Training data selection} section).

\subsection{Similar Crystallization Pathways in Disparate Systems} \label{subsec:paths}

\begin{figure*}
    \includegraphics[width=1\linewidth]{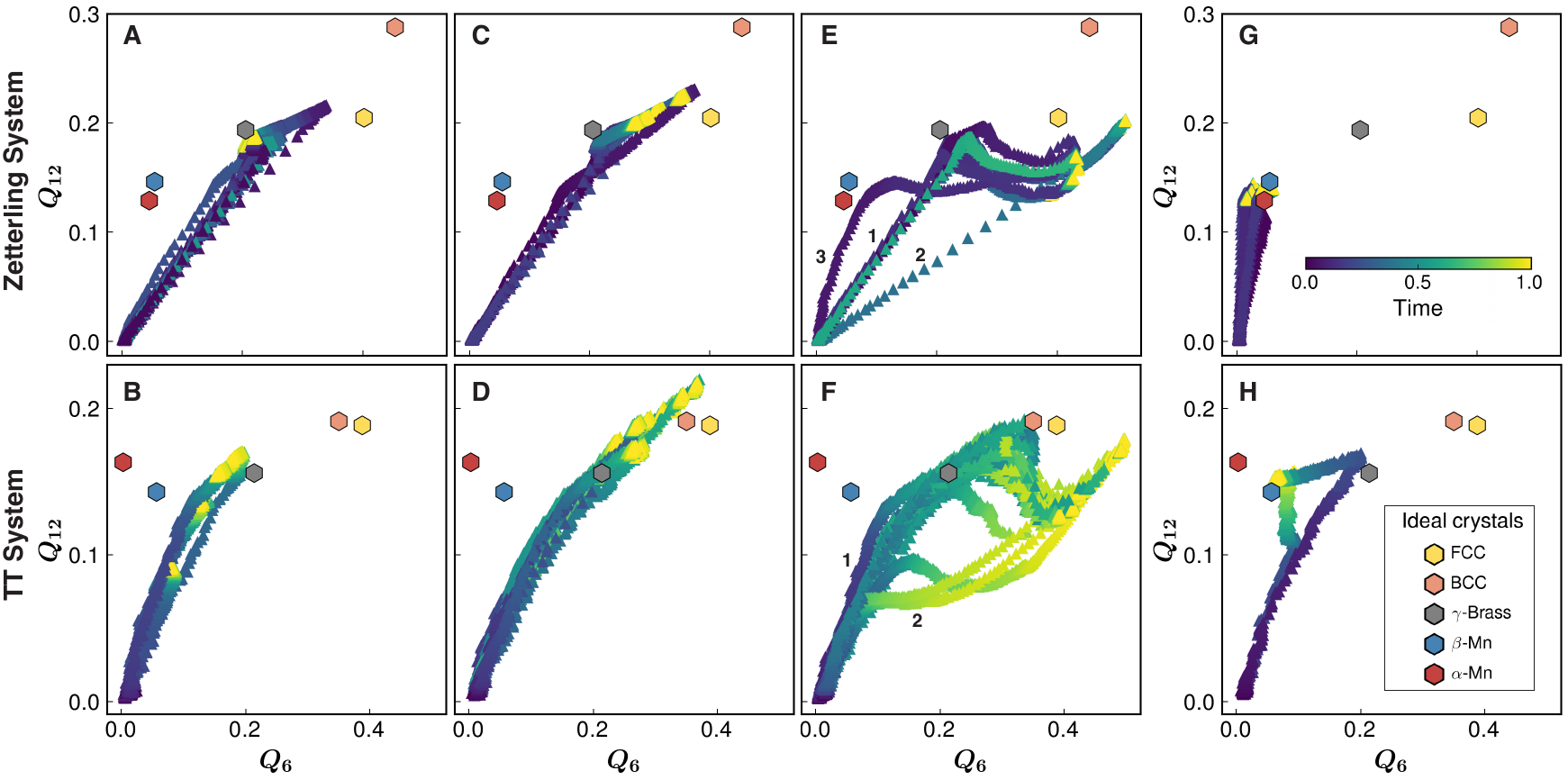}
    \caption{Crystallization pathways of the Zetterling system (A, C, E, G) and the TT system (B, D, F, H).
    Each triangle denotes a configuration along the trajectory (quantified by Steinhardt order parameters~\cite{steinhardt1983bond} $Q_{6}$ and $Q_{12}$) and the color indicates timestep. (The time span for each trajectory is scaled to a range from $0$ to $1$, and the lengths of the trajectories vary.) Thermalized ideal crystals represented with hexagons are also shown for reference. The two systems share the following common pathways leading to: (A, B) $\gamma$-Brass, (C, D) mixtures of $\gamma$-Brass and BCC, and (E, F) FCC. Both systems have more than one type of FCC pathway, as indicated by the numbers in (E, F). In addition, the pathways to (G) $\beta$-Mn and (H) $\alpha$-Mn are only observed in the Zetterling system and the TT system, respectively.
   }
    \label{fig:pathways}
\end{figure*}

In addition to isopolymorphism, the crystallization pathways of the two systems also share many similarities in the temporal order in which the crystal polymorphs form. First, both systems almost always first form large $\gamma$-Brass grains, even when they eventually form other crystals (Figs.~\ref{fig:pathways}A-F). As a result, the transition from fluid to BCC is a two-step process for both systems: fluid $\rightarrow$ $\gamma$-Brass followed by a solid-solid transition from $\gamma$-Brass $\rightarrow$ BCC.

Notably, when both systems form FCC, they each exhibit two types of pathways (Figs.~\ref{fig:pathways} E and F): pathways in which the system forms large $\gamma$-Brass and BCC crystal grains before forming FCC (Pathway $1$), following Ostwald's rule of stages~\cite{ostwald1897studien}, and pathways in which the system forms FCC crystal grains directly from the fluid (Pathway $2$). For the Zetterling system, we also observed a third type of pathway in which the system forms both large $\gamma$-Brass and $\beta$-Mn crystal grains before forming FCC (Pathway $3$). As will be discussed below, along Pathway $3$ the Zetterling system happened to form $\gamma$-Brass and $\beta$-Mn crystal grains at the same time, yet the $\beta$-Mn LSs play no role in the formation of FCC. Therefore, this pathway is essentially the same as Pathway $1$.

When the Zetterling system forms $\beta$-Mn and the TT system forms $\alpha$-Mn, both of which are also complex crystals, structural order also emerges directly from the fluid phase (Figs.~\ref{fig:pathways} G and H). However, for the TT system, the only two $\alpha$-Mn pathways we sampled in our unbiased simulations appear to be different, whereas the Zetterling system follows the same path when forming $\beta$-Mn.

Using the supervised classification approach detailed in the \hyperref[subsec:classify]{Local Structure Classification} section, we dissected these crystallization pathways down to the single-particle level. With our LS analysis (\hyperref[method: ls_analysis]{Local Structure Evolution Analysis}), we found even more shared traits in the local structural evolution of the two systems beyond the same multistep pathways discussed here, which we present in the section below. 

\subsection{Common Transition 1: \texorpdfstring{Fluid $\rightarrow \gamma$-Brass $\rightarrow$ BCC}{Fluid → γ-Brass → BCC}} \label{subsec:toBrass}

\begin{figure*}[t]
    \centering
    \includegraphics[width=1\linewidth]{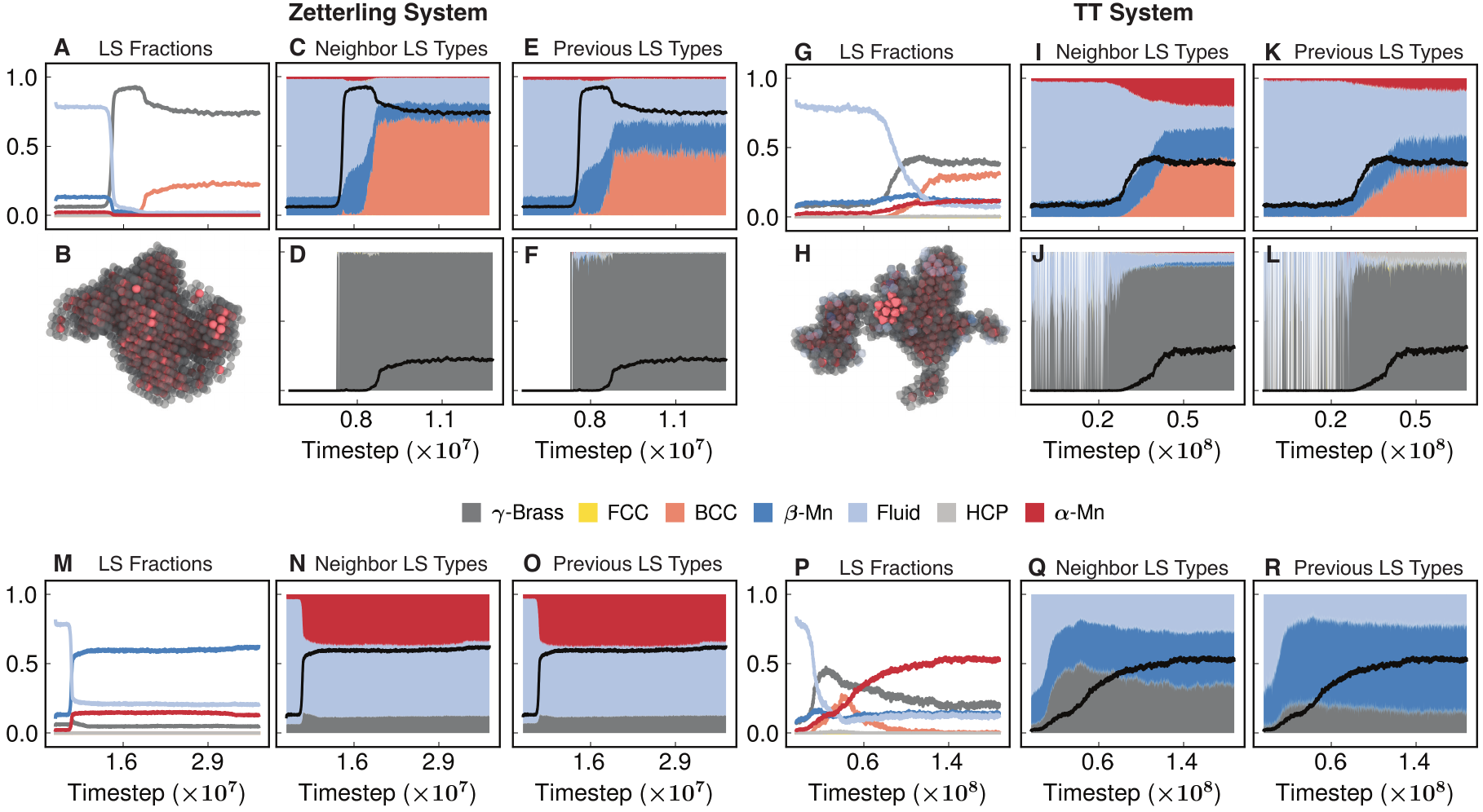}
    \caption{LS analysis of two exemplary trajectories of the Zetterling system in which the system formed mixtures of BCC and $\gamma$-Brass (A-F) and $\beta$-Mn (M-O), and LS analysis of two exemplary trajectories of the TT system in which the system formed mixtures of BCC and $\gamma$-Brass (G-L) and $\alpha$-Mn (P-R). \textbf{(A, G, M, P)} LS fraction evolution along the trajectories. \textbf{(C, D, I, J)} LS types of the \textit{nearest neighbors} of the $\gamma$-Brass-like particles (C, I) and the BCC-like particles (D, J). \textbf{(E, F, K, L)} LS types of the \textit{particles that transition to} $\gamma$-Brass-like particles (E, K) and BCC-like particles (F, L). \textbf{(B, H)} BCC clusters surrounded by $\gamma$-Brass particles. $\gamma$-Brass-like particles are rendered transparent for better visualization of the BCC core. \textbf{(N, Q)} LS types of the \textit{nearest neighbors} of the $\beta$-Mn-like particles (N) and the $\alpha$-Mn-like particles (Q). \textbf{(O, R)} LS types of the \textit{particles that transition to} $\beta$-Mn-like particles (O) and $\alpha$-Mn-like particles (R). In the LS analysis plots, the LS fractions of the particle types of interest ($\gamma$-Brass and BCC in A-L; $\beta$-Mn and $\alpha$-Mn in M-R) are drawn as black lines, and the percentages of the LSs are shown as stacked color blocks, whose height at each timestep sums up to 1.}
    \label{fig:brass-bcc-mn}
\end{figure*}

$\gamma$-Brass forms from the fluid phase in both the Zetterling system and the TT system, and it is usually the first crystal to form even when the system forms a different crystal eventually. Two example trajectories for each system are shown in Figs.~\ref{fig:brass-bcc-mn} A and G.

Through our analysis of the LSs (detailed in the \hyperref[method: ls_analysis]{Local Structure Evolution Analysis} section), we found that for both systems, during the growth stage of $\gamma$-Brass clusters (timestep $\leq 0.8 \times 10^{7}$ for the Zetterling system and timestep $\leq 0.4 \times 10^{8}$ for the TT system), the nearest neighbors of $\gamma$-Brass particles have fluid and $\beta$-Mn LSs, and more neighbors exhibit fluid LSs (Figs.~\ref{fig:brass-bcc-mn} C and I). In other words, $\gamma$-Brass clusters and individual $\gamma$-Brass-like particles are mostly surrounded by fluid-like particles in the pre-nucleation fluid. Additionally, we found that most of the particles that transition to $\gamma$-Brass LSs were fluid-like particles (Fig.~\ref{fig:brass-bcc-mn} E and K), which corroborates on the single-particle level that both systems develop $\gamma$-Brass order directly from the fluid phase.

When both systems form $\gamma$-Brass, in addition to the $\gamma$-Brass LSs, they also almost always develop BCC LSs.
More specifically, for the Zetterling system, $\gamma$-Brass LSs almost always develop first, which promotes the formation of BCC LSs. For the TT system, $\gamma$-Brass LSs develop in all pathways, regardless of the final equilibrium crystal structure, and thus BCC LSs always appear.

To assess the relative onset and evolution of these two types of LSs, we plotted the fraction of $\gamma$-Brass LSs directly against the fraction of BCC LSs for all the relevant pathways (Fig.~\ref{fig:zetterling_bcc_brass} for the Zetterling system and Fig.~\ref{fig:tt_bcc_brass} for the TT system). For the Zetterling system, BCC-like particles start to appear when at least $10 \% - 20 \%$ of the particles are already $\gamma$-Brass-like, and for most of the pathways, especially the ones leading to $\gamma$-Brass, this number is even higher (above $60 \%$) (Fig.~\ref{fig:zetterling_bcc_brass}A). In such cases, the system has already formed large $\gamma$-Brass grains. For the TT system, just like in the Zetterling system, BCC LSs start to appear when the fraction of $\gamma$-Brass-like particles in the system is at least around $20 \%$ (Fig.~\ref{fig:tt_bcc_brass}).
 
In other words, BCC LSs develop only when $\gamma$-Brass LSs are present and they grow at a slower rate than the $\gamma$-Brass LSs. The maximum amount of $\gamma$-Brass LSs also limits the largest size of the BCC crystal grains. Sometimes the BCC-like particles are able to form sizable crystal grains, resulting in mixtures of $\gamma$-Brass crystal grains and BCC grains (Figs. S1B and S2B).

In both systems, for the pathways that lead to large grains of BCC crystals, we found that the BCC clusters are always surrounded by $\gamma$-Brass-like particles. For the Zetterling system, nearly $100 \%$ of the nearest neighbors of BCC clusters are $\gamma$-Brass-like (Fig.~\ref{fig:brass-bcc-mn}D), and the percentage of $\gamma$-Brass-like neighbors is over $90 \%$ for the TT system (Fig.~\ref{fig:brass-bcc-mn}J). Examples of BCC clusters surrounded by $\gamma$-Brass-like particles are shown in Fig.~\ref{fig:brass-bcc-mn} B and H. Our LS transition analysis further reveals that as the BCC clusters grow (timestep between $0.8 \times 10^{7}$ and $0.9 \times 10^{7}$ in the Zetterling system, and timestep between $0.2 \times 10^{8}$ and $0.5 \times 10^{8}$ for the TT system), most of the particles that adopt BCC LSs are previously $\gamma$-Brass-like (Figs.~\ref{fig:brass-bcc-mn} F and L). These local structure analyses demonstrate that the $\gamma$-Brass $\rightarrow$ BCC transition occurs within $\gamma$-Brass crystal grains.

\subsection{Common Transition 2: \texorpdfstring{Fluid $\rightarrow$ FCC}{Fluid → FCC}} \label{subsec:toFCC}

As mentioned in the \hyperref[subsec:paths]{Similar Crystallization Pathways in Disparate Systems} section, both the Zetterling system and the TT system exhibit the following two types of pathways when forming FCC: (1) Pathway $1$ (more common), along which the system forms $\gamma$-Brass and BCC crystal grains before the system forms FCC and (2) Pathway $2$, in which FCC forms directly from the fluid. Additionally, we observed a very rare pathway for the Zetterling system, Pathway $3$, in which the system forms both $\gamma$-Brass and $\beta$-Mn crystal grains before it forms FCC.

\begin{figure*}[t!]
    \centering
    \includegraphics[width=0.85\linewidth]{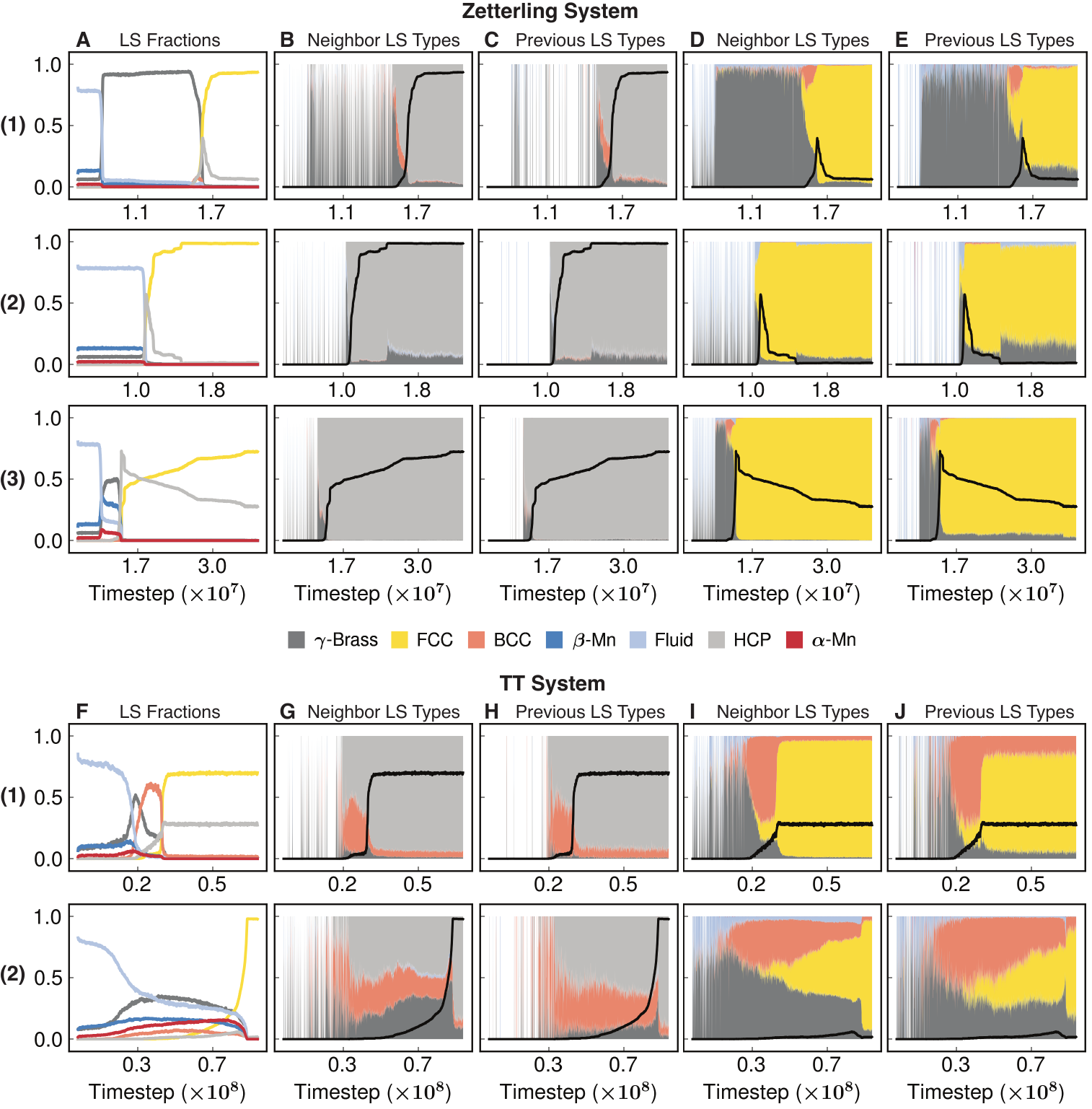}
    \caption{LS analysis of five exemplary trajectories of the Zetterling system (A-E) and the TT system (F-J) in which the system formed FCC. \textbf{(A, F)} LS fraction evolution along the trajectories for the Zetterling system ((1) Pathway $1$, (2) Pathway $2$, and (3) Pathway $3$) and the TT system ((1) Pathway $1$ and (2) Pathway $2$). \textbf{(B, D, G, I)} LS types of the \textit{nearest neighbors} of the FCC-like particles (B, G) and the HCP-like particles (D, I). \textbf{(C, E, H, J)} LS types of the \textit{particles that transition to} FCC-like particles (C, H) and HCP-like particles (E, J). In the LS analysis plots, the LS fractions of the particle types of interest (FCC and HCP) are drawn as black lines, and the percentages of the LSs are shown as stacked color blocks, whose height at each timestep sums up to 1.
   }
    \label{fig:fcc}
\end{figure*}

For Pathway $1$ of the Zetterling system, both the FCC and HCP LSs transition predominantly from $\gamma$-Brass LSs at the early growth stage of FCC (the height of the gray $\gamma$-Brass block is at least $0.5$ at the first elbow points of the black lines in Fig.~\ref{fig:fcc}(1)B-E). At this stage, some FCC and HCP particles also transition from BCC LSs. As the number of FCC-like particles increases, $\gamma$-Brass and BCC LSs rapidly disappear, and HCP LSs become the dominant source for FCC-like particles. As shown in Fig.~\ref{fig:fcc}(3)B-E), the LS transitions along Pathway $3$ are very similar to Pathway $1$. Indeed, after identifying the LS types for each particle, we directly observed this particular trajectory, and found that the $\beta$-Mn- and $\gamma$-Brass-like particles grow into larger crystal clusters simultaneously, but FCC forms from the $\gamma$-Brass crystal grain. Therefore, the seemingly different Pathway $3$ of the Zetterling system is, in fact, the same as Pathway $1$, along which the system forms FCC via forming $\gamma$-Brass first. Specifically, we note that the $\beta$-Mn LSs do not participate in the formation of FCC, even though the Zetterling system also forms $\beta$-Mn grains in Pathway $3$ (Fig.~\ref{fig:pathways}E).

For Pathway $2$, there is only a very small amount of $\gamma$-Brass LSs in the pre-nucleation fluid of the system, and $\gamma$-Brass LSs become HCP and FCC LSs when the system crystallizes. At the same time, some fluid LSs become HCP LSs, which then become FCC LSs (Fig.~\ref{fig:fcc}(2)B-E).

Similar to Pathway $1$ of the Zetterling system, in \textit{both} Pathways $1$ and $2$ of the TT system, $\gamma$-Brass and BCC LSs become HCP and FCC LSs. Once the FCC LSs start to develop, fractions of $\gamma$-Brass- and BCC-like particles decrease immediately, and HCP LSs become the primary source for FCC LSs. Due to the higher fraction of BCC particles in the TT system along these pathways, FCC particles have a higher percentage of BCC-like neighbors and precursors, compared to the Pathway $1$ of the Zetterling system.

A summary of the local structure transitions is presented in Fig.~\ref{fig:local_pathways}B.

\subsection{Crystallization pathways of \texorpdfstring{$\beta$-Mn/$\alpha$-Mn}{β-Mn/α-Mn}} \label{subsec:toMn}

In addition to $\gamma$-Brass, the Zetterling system can also form $\beta$-Mn (Fig.~\ref{fig:polymorphism}C) and the TT system can form $\alpha$-Mn (Fig.~\ref{fig:polymorphism}L), both of which are also complex crystals.

As can be seen from Fig.~\ref{fig:pathways}G, in the Zetterling system, $\beta$-Mn forms directly from the fluid phase.
Therefore, the pathways leading to $\beta$-Mn are the few exceptions in which the system does not form $\gamma$-Brass crystal grains, and thus almost no BCC LSs are present along these pathways. Our LS analysis confirms that the fluid LSs become $\beta$-Mn LSs in these systems (Fig.~\ref{fig:brass-bcc-mn}O). As shown in Fig.~\ref{fig:brass-bcc-mn}N, over $90 \%$ of the nearest neighbors of the $\beta$-Mn-like particles have fluid-like LSs during the growth stage (timestep $\leq 1 \times 10^{7}$).

The TT system assembles into large grains of $\alpha$-Mn very rarely (Fig.~\ref{fig:polymorphism}F), and the only two trajectories are shown in Fig.~\ref{fig:pathways}H. Interestingly, even though the entire system exhibits $\alpha$-Mn order (as evidenced by the BOD in Fig.~\ref{fig:polymorphism}L), the $\text{Q}_{6}$ and $\text{Q}_{12}$ values of the assembled $\alpha$-Mn are closer to those of the reference $\beta$-Mn crystal; yet upon visual inspection of the crystal we found no $\beta$-Mn crystal grains. We confirmed that in these systems the $\alpha$-Mn LSs are the most dominant LSs, and the deviation of the system from ideal $\alpha$-Mn crystal is because of the presence of $\gamma$-Brass, $\beta$-Mn and fluid LSs (Fig.~\ref{fig:brass-bcc-mn}P)

The two trajectories leading to $\alpha$-Mn appear to be very different from one another (Fig.~\ref{fig:pathways}H), and this is because the TT system first formed a $\gamma$-Brass crystal grain in one trajectory (LS fraction is shown in Fig.~\ref{fig:brass-bcc-mn}P), whereas it only formed $\alpha$-Mn in the other one. For these two trajectories, we observed no significant difference in the LS types of the nearest neighbors of the $\alpha$-Mn-like particles, or the LS types of the particles that become $\alpha$-Mn-like.
In both cases, most of the particles involved have $\gamma$-Brass, $\beta$-Mn, and fluid LSs.

\subsection{Origin of Similarities}

\begin{figure*}[t!]
    \centering
    \includegraphics[width=0.95\linewidth]{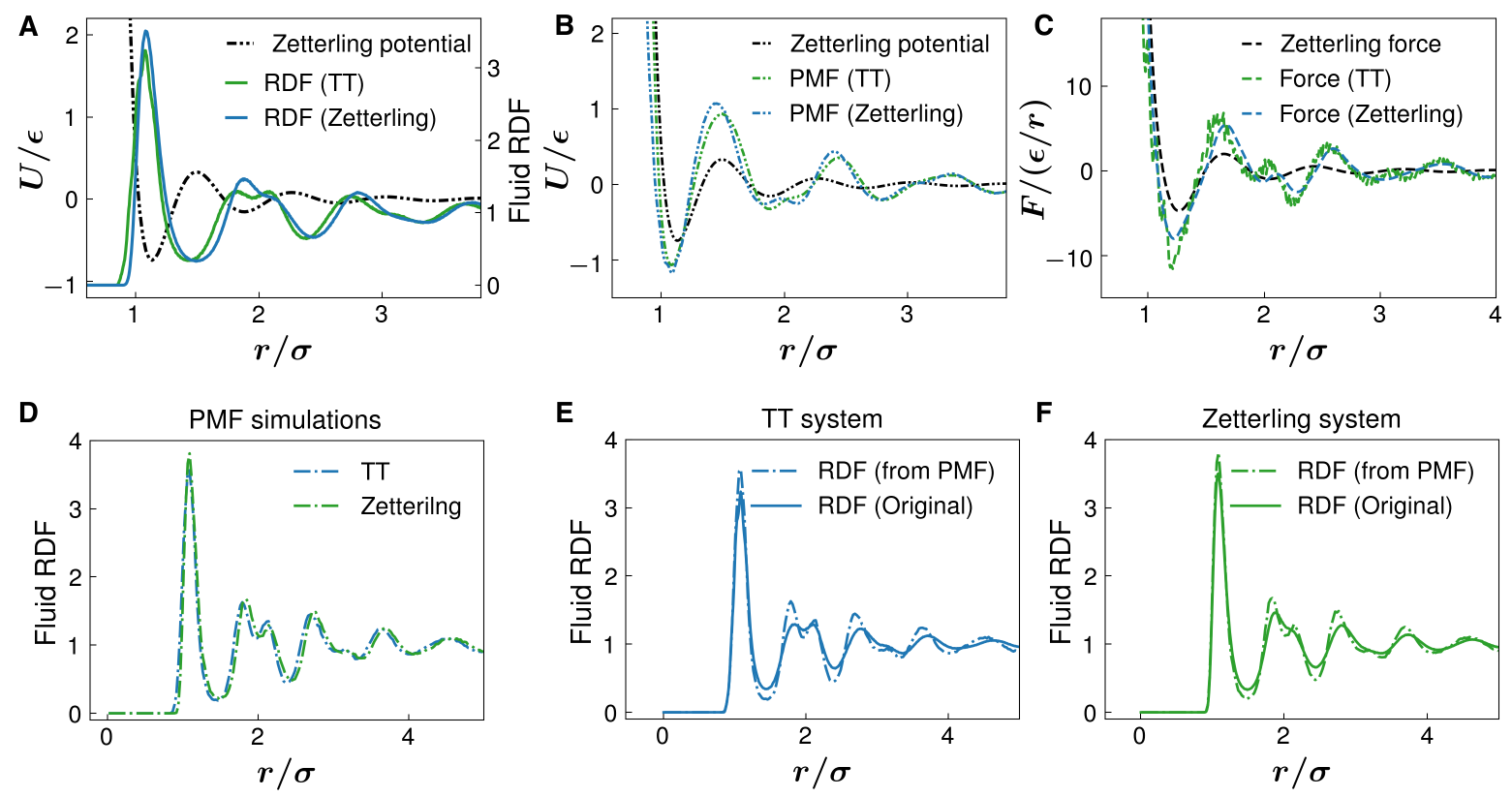}
    \caption{
    (A) RDF profiles of the fluid phase computed from the self-assembly simulations for both systems. The Zetterling potential is shown in comparison. (B) The PMFs we calculated using the RDF in (A), shown in comparison with the Zetterling potential. (C) Forces we calculated by taking the gradient of the PMFs in (B), shown in comparison with the Zetterling force. (D) The fluid RDF profiles computed from simulations performed with the PMF and force in (B) and (C). We also compare these RDF profiles with those that correspond to the original (E) TT shape and (F) the Zetterling potential.
   }
    \label{fig:pmf_analysis}
\end{figure*}

The similarities we report here for the very different Zetterling and TT systems are striking. To determine whether these similarities extend beyond the crystallization pathways, we computed the radial distribution functions (RDFs) in the initial fluid phases for each system prior to crystallization.  Fig.~\ref{fig:pmf_analysis}A shows that the peaks in the two RDFs align at the same $r/\sigma$ values despite the distinct origins of the particle interactions. Also, both RDFs have a split second peak, commonly observed in glassy systems and amorphous metals and alloys~\cite{bernal1964bakerian, finney1977modelling,ding2015second}.

To rationalize this striking similarity in the fluid structure, which potentially gives rise to the isopolymorphism and similarities in the crystallization pathways, we computed the potentials of mean force (PMFs) for each system from the RDFs in Fig.~\ref{fig:pmf_analysis}A. As shown in Fig.~\ref{fig:pmf_analysis}B, the PMFs of the two systems are highly similar. For each system, we performed an MD simulation of a new system of point particles interacting via a pair potential given by the PMF and the associated force (Fig.~\ref{fig:pmf_analysis}C) obtained by taking the gradient of the PMF (details can be found in the \hyperref[method:pmf]{Potential of Mean Force Analysis} section). The RDF profiles computed from these simulations almost completely overlap (Fig.~\ref{fig:pmf_analysis}D). Moreover, the peaks of the RDFs align with those of the RDFs of the original systems (Figs.~\ref{fig:pmf_analysis} E and F). 

These results suggest that we can represent the entropic interactions in the TT system with a two-body potential, and that this potential is effectively equivalent to the PMF of the Zetterling system at the state point of interest. Importantly, this PMF -- which is state point dependent -- is not identical to the Zetterling potential, but produces the same local structure as the Zetterling potential prior to crystallization. This means that the statistical, emergent entropic interactions in the TT system at our thermodynamic state point of interest can be described in terms of an effective, local interaction given by the effective potential (PMF) describing the Zetterling system; it is this mapping that gives rise to the similar local structures prior to crystallization and the subsequent similar crystallization pathways.

\begin{figure*}[t!]
    \centering
    \includegraphics[width=0.95\linewidth]{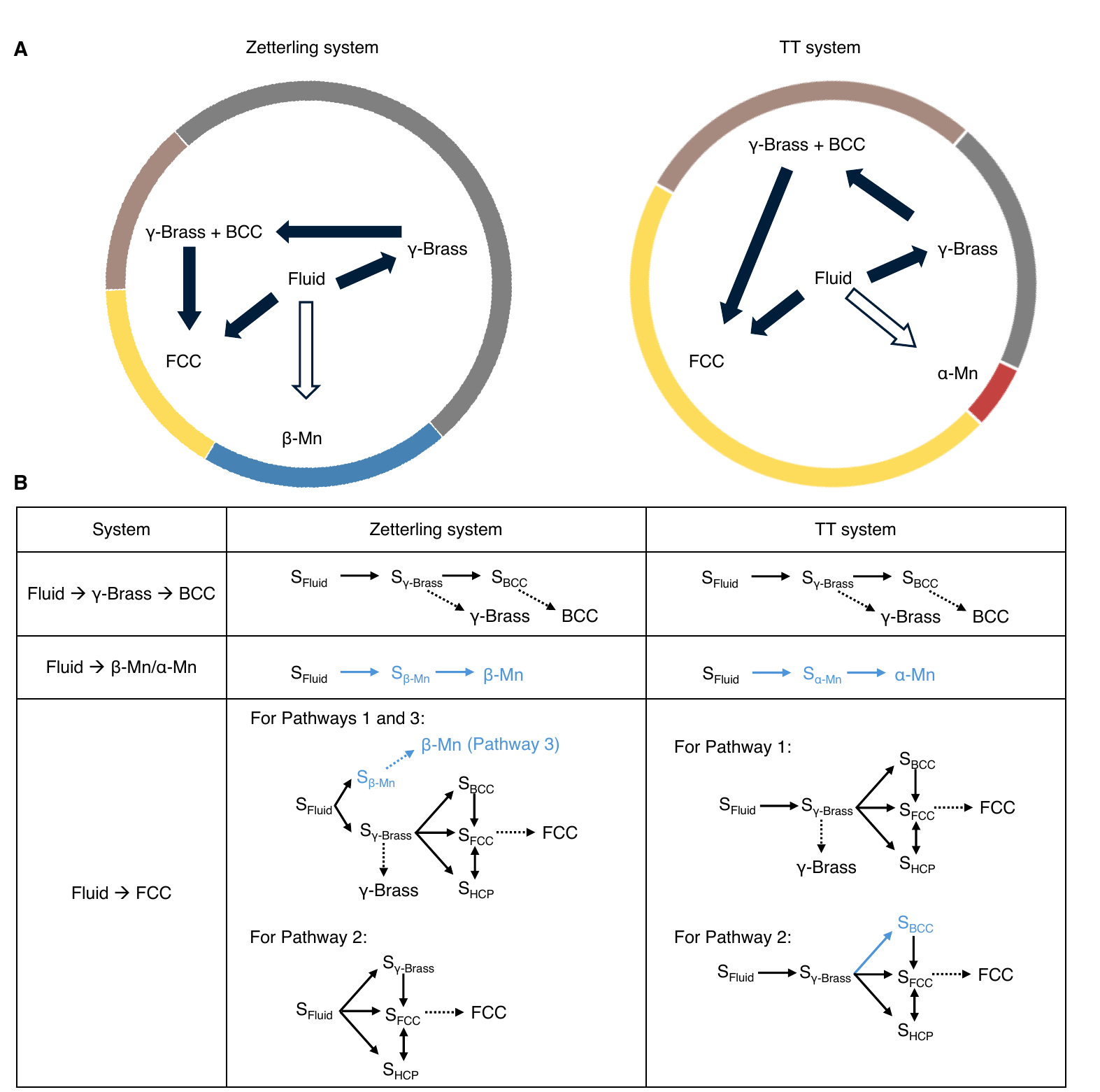}
    \caption{
    (A) Frequencies of the observed crystal structures and a summary of the transition pathways. Solid arrows represent transitions common to both systems, while hollow arrows indicate transitions unique to one system. (B) Comparison of the roles of local structures (LSs, denoted as "S") along the transition pathways shown in the top panel. Dashed arrows point to the final crystal structures formed by the entire system. Differences in local transition pathways are highlighted in blue.
   }
    \label{fig:local_pathways}
\end{figure*}

\section*{Discussion} \label{conclusion}

In summary, we performed extensive molecular dynamics simulations for two complex-crystal-forming systems with different particle interaction origins at thermodynamic state points at which both systems have been reported to form a common complex crystal, $\gamma$-Brass. We found that the two systems can crystallize into four competing polymorphs, three of which are the same: BCC, FCC and $\gamma$-Brass, and the fourth polymorphs are allotropes of the manganese metal ($\beta$-Mn and $\alpha$-Mn). Moreover, the two systems follow the same multistep transition pathways among the shared polymorphs. The schematics summarizing the pathways are presented in Figs.~\ref{fig:local_pathways} A and B.

Notably, by analyzing the local structures and tracking their evolution, we uncovered microscopic mechanisms that may not be detected with global metrics. For example, the BCC phase nucleates and grows from within a $\gamma$-Brass phase. We believe the considerations detailed in the \hyperref[subsec:classify]{Local Structure Classification} section, which are broadly applicable to investigations of local structures along crystallization pathways involving complex crystals, will provide guidance on using ML methods for such studies.

Additionally, the almost identical match among the fluid RDF profiles of the Zetterling and TT systems and those obtained from the potentials of mean force suggests that the interactions in the TT system, which are purely entropic in nature, can be mapped to a pair potential that is equivalent to the PMF of the Zetterling system. This similarity in the particle interactions explains the isopolymorphism and the similar crystallization pathways observed for these two systems.

\section{Methods}\label{methods}

\subsection{Simulation Protocols}\label{method:sim}

\subsubsection{Zetterling System}
To corroborate the crystallization of $\gamma$-Brass reported by Zetterling et al.~\cite{zetterling2000gamma} and to generate our own data, we performed molecular dynamics simulations of point particles interacting with the oscillatory pair potential as defined in Ref.~\citenum{zetterling2000gamma}, which is truncated and shifted to zero at its third minimum. Simulations of $N = 80, 000$ particles were run in the NPT ensemble at $k_{B}T = 0.85\epsilon$ and $P = 6.5\epsilon / {\sigma}^{3}$ with the open-source simulation package HOOMD-blue~\cite{anderson2020hoomd} (versions above 2.9.7). Pressure and temperature were maintained with the MTK barostat~\cite{martyna1994constant} and the Nos\'{e}-Hoover thermostat~\cite{hoover1985canonical}. For these simulations, we used a timestep of $dt=0.001\tau$, where $\tau = \sigma(m/\epsilon)^{1/2}$, and $m$ is particle mass (set to 1). To generate random initial configurations, we initialized the system in a sparse cubic array in a cubic box at a low density ($\rho\sigma^{3} = 0.3$), and thermalized the system in the canonical ensemble (NVT) at an elevated temperature $k_{B}T = 1.0\epsilon$ for $3 \times 10^{6}$ timesteps. Then the system was compressed to the target pressure and was further thermalized at $k_{B}T = 1.0\epsilon$ for another  $3 \times 10^{6}$ timesteps in the NPT ensemble, during which we enforced a cubic box constraint on the simulation box while the system remained disordered. We subsequently decreased the temperature to $k_{B}T = 0.85\epsilon$, relaxed the cubic box constraint to only enforcing orthorhombic box symmetry, and equilibrated the system for at least $5 \times 10^{7}$ timesteps. Once the temperature was decreased, crystallization usually occurred spontaneously after at least several million timesteps.

\subsubsection{Truncated Tetrahedron System}
To study the self-assembly pathways of the TT system, we ran molecular dynamics simulations of truncated tetrahedra (truncation parameters $(\alpha_{a}, \alpha_{c}) = (0.25, 0.5)$~\cite{teich2019identity}) that interact with the anisotropic Weeks-Chandler-Anderson pair potential (AWCA)~\cite{ramasubramani2020mean}. All simulations were also performed with HOOMD-blue (versions above $4.0.0$), and the vertices of the shape were generated with the Python library Coxeter~\cite{ramasubramani2021coxeter} and then scaled by a factor of $2$. We used the following parameters for the potential: $\epsilon = 0.1$, $\alpha = 0$ for repulsive contact-contact and center-center interaction, and $\sigma_{i} = \sigma_{i} = \text{d}$, where $\text{d}$ is the insphere diameter of the truncated tetrahedron. Simulations of $N = 10,000$ truncated tetrahedra were run in the NPT ensemble at $k_{B}T = 1\epsilon$ and $P = 0.32\epsilon / {\sigma}^{3}$ with a timestep of $dt=0.005\tau$. Pressure and temperature were maintained with the MTK barostat and the Bussi-Donadio-Parrinello thermostat~\cite{bussi2007canonical}. To generate random initial configurations, we initialized the system in a sparse cubic array in a cubic simulation box, and thermalized the system at a much lower pressure ($0.05\epsilon / {\sigma}^{3}$) for $1 \times 10^{6}$ timesteps. The system was subsequently compressed over $1 \times 10^{6}$ timesteps to reach the target pressure, during which we enforced a cubic box constraint while the system remained disordered. At the target pressure, we relaxed the cubic box constraint to only enforcing orthorhombic box symmetry, and the system was equilibrated for at least $5 \times 10^{7}$ timesteps. Crystallization usually occurred spontaneously after at least $1 \times 10^{7}$ timesteps.

\subsection{Crystal Structure Identification}\label{method:global}
To determine the structure of the bulk crystals of the two systems, we directly inspected the structures, and computed the following structure metrics for both the self-assembled structures and the thermalized ideal crystals using the open-source analysis package \textit{freud}~\cite{ramasubramani2020freud}: (1) radial distribution function, (2) 1D structure factor, (3) Steinhardt order parameters~\cite{steinhardt1983bond}, (4) diffraction pattern and (5) bond-orientational order diagram (BOD). We then compared the results of the self-assembled structures against those of the thermalized ideal crystals as well as previously published data if any was available.

To best showcase the pathways of both systems in terms of system-wide order parameters (Fig.~\ref{fig:pathways}), we used two different types of neighbor lists to compute the global Steinhardt order parameters $\text{Q}_{6}$ and $\text{Q}_{12}$ for each frame of the trajectories. We used the Voronoi neighbor list for computing $\text{Q}_{6}$, and the first two coordination shells (determined from the radial distribution function) for $\text{Q}_{12}$. For the Zetterling system, the radius cutoff that includes the first two coordination shells is $2.5\sigma$, and for the TT system the cutoff is $11.5\sigma$.

\subsection{Local Structure Features} \label{method:features}

To ensure our supervised classifiers are informed of the thermal noise and defects intrinsic to the self-assembled crystals as well as the thermodynamic (meta-)stability of the polymorphs, we use crystals thermalized at/near the self-assembly simulation state points in our stability tests as the training data, instead of perfect crystals with zero thermal noise or defects. We discuss how we selected the training data in the \textit{SI Appendix Training data selection} section.

We adopt the approach of Ref.~\cite{dice2021complex} to compute the following local structure order parameters as features using \textit{freud}: (1) the Minkowski Structure Metrics~\cite{mickel2013shortcomings} (MSMs) $q'_{\ell}$ and its neighbor-averaged variant $\bar{q'_{\ell}}$, (2) third-order rotationally invariant Steinhardt order parameter $w_{\ell}$ and its neighbor-averaged variant $\bar{w_{\ell}}$, and (3) the Voronoi density (inverse of Voronoi volume). The MSMs are Steinhardt order parameters ($q_{\ell}$) computed with weighted neighbor lists constructed via Voronoi tessellation. We use the MSMs because the Voronoi tessellation neighbor finding approach is agnostic to simulation volume/pressure, or the system itself, and it gives rise to a continuous neighbor list.

\subsection{Supervised Classification Method}\label{method: supervised-classify}

We performed the classification of LSs using the machine learning package scikit-learn~\cite{pedregosa2011scikit}. As mentioned in the \hyperref[subsec:classify]{Local Structure Classification} section, we used two $k$-Nearest Neighbor classifiers to first classify fluid- vs crystal-like particles and then identify the LS types of the crystal-like particles.
The $k$ values we used are reported in Table \ref{tab:k_values}.

\begin{table}[h]
\centering
\caption{$k$ values used for the fluid/crystal classifier and crystal LS classifier for the Zetterling system and the TT system.}
\label{tab:k_values}
\begin{tabular}{ccc}
\hline
System& Zetterling& TT\\ 
\hline
Fluid/crystal classifier& 11& 15\\ 
Crystal LS classifier& 55& 55\\
\hline
\end{tabular}
\end{table}

\subsection{Local Structure Evolution Analysis}
\label{method: ls_analysis}
After classifying the LS type for each individual particle along the crystallization pathways, we further analyzed the local structure transition in the following two aspects: (1) We checked the LS types of the nearest neighbors of the particles/clusters of interest. For example, we checked the LS types of the nearest neighbors of BCC-like particles (Fig.~\ref{fig:brass-bcc-mn}B and H). When the BCC-like particles form crystal grains, the nearest neighbors typically form a layer that encompasses the grain. In this case, the nearest neighbors are mostly of the $\gamma$-Brass type.(2) We also tracked the transitions in the LS types. In other words, for all the particles that have the LS type of interest at a given timestep, we check their LS types at a previous timestep. With this, we are able to confirm the types of particles that become e.g., BCC-like, and thus determine if a type of LS is the precursor of another type.

\subsection{Potential of Mean Force Analysis}\label{method:pmf}
To directly compare the interactions of the two systems, we first computed the radial distribution function (RDF) of the fluid phases. For each system, we used the first two frames of three independent assembly simulation trajectories obtained at the target thermodynamic state points, in which the system was still in the fluid phase. Then we computed the potential of mean force (PMF) using the following equation:
\begin{align}
    \begin{aligned}
        V_{\text{PMF}} = -k_BT \ln g(r),
        \label{eq:pmf}
    \end{aligned}
\end{align}
where $k_BT$ is the temperature at which we performed the simulations and $g(r)$ is the RDF calculated in the previous step. We then fitted a one dimensional spline to the $V_{\text{PMF}}$ values using Scipy~\cite{2020SciPy-NMeth}'s \texttt{UnivariateSpline} function (\texttt{k} $= 3$ and \texttt{s} $= 0.1$ in both cases). As $g(r) = 0$ for small $r$, we manually set the $V_{\text{PMF}}$ values at these $r$ values to be large numbers such that the fitted potential has a hard core at where $g(r) = 0$. We then calculated the force by taking the gradient of the $V_{\text{PMF}}$ spline function using Scipy. With the potential energy values ($V_{\text{PMF}}$) and the associated force, we performed molecular dynamics simulations using HOOMD-blue. For each system, we used $N = 4,000$ point particles, and followed the same steps detailed in the \hyperref[method:sim]{Simulation Protocols} section to generate random initial configurations at the target thermodynamic state points. 

\section*{Acknowledgements}

This research was supported by a Computational and Data-Enabled Science and Engineering (CDS\&E) grant from the U.S. National Science Foundation, Division of Materials Research Award No. DMR 2302470. This work used NCSA Delta, SDSC Expanse and Purdue Anvil through allocation DMR 140129 from the Advanced Cyberinfrastructure Coordination Ecosystem: Services \& Support (ACCESS) program, which is supported by National Science Foundation grants No. 2138259, No. 2138286, No. 2138307, No. 2137603, and No. 2138296. Computational resources and services were also provided by Advanced Research Computing at the University of Michigan, Ann Arbor. The authors also thank Dr. Joshua A. Anderson, Dr. Sun-Ting Tsai and Dr. R. Allen LaCour for their helpful suggestions and insights.

\section*{Competing interests}
The authors declare no competing interests

\section*{Data availability}

The data that support the findings of this study are available from the corresponding author upon reasonable request.

\section*{Code availability}
The code used to perform the simulations and analyze the results of this study are available from the corresponding author upon reasonable request.

\section*{Author contribution}
C.S.Z., D.F., and S.C.G. designed research; C.S.Z., D.F., and S.C.G. performed research; C.S.Z. curated simulation data; C.S.Z. and D.F. analyzed data; S.C.G. supervised research; and C.S.Z., D.F., and S.C.G. wrote the paper.





\bigskip






\bibliography{sn-bibliography}

\makeatletter
\@input{supplementary_aux.tex}
\makeatother
\end{document}


\beginsupplement

\title[Article Title]{Supporting Information for Anatomy of a Complex Crystallization Pathway}

\author[1]{\fnm{Charlotte Shiqi} \sur{Zhao}}\email{zshiqi@umich.edu}

\author[1]{\fnm{Domagoj} \sur{Fijan}}\email{dfijan@umich.edu}

\author*[1,2]{\fnm{Sharon} \sur{C. Glotzer}}\email{sglotzer@umich.edu}

\affil[1]{\orgdiv{Department of Chemical Engineering}, \orgname{University of Michigan}, \orgaddress{\street{2800 Plymouth Road, Building 10}, \city{Ann Arbor}, \postcode{48109}, \state{Michigan}, \country{United States}}}

\affil[2]{\orgdiv{Biointerfaces Institute}, \orgname{University of Michigan}, \orgaddress{\street{2800 Plymouth Road, Building 10, A175}, \city{Ann Arbor}, \postcode{48109}, \state{Michigan}, \country{United States}}}



\maketitle
\pagebreak

\tableofcontents
\pagebreak

\addcontentsline{toc}{section}{Supplementary Notes}
\section*{Supplementary Notes}

\addcontentsline{toc}{subsection}{Supplementary Note 1 – Crystal Stability Test}
\subsection*{Supplementary Note 1 – Crystal Stability Test}\label{method:stability}

\addcontentsline{toc}{subsubsection}{Zetterling System}
\subsubsection*{Zetterling System}
For the Zetterling system, all of the crystal polymorphs ($\gamma$-Brass, BCC, FCC, $\beta$-Mn, $\alpha$-Mn) can remain mechanically stable at the self-assembly simulation condition ($P = 6.5\epsilon / {\sigma}^{3}$, $k_{B}T = 0.85\epsilon$).
For these simulations, we initialized the Zetterling system in the crystal polymorphs at $k_{B}T = 0.85\epsilon$ and densities shown in Table \ref{tab:zetterling_stability_result}.
We then ran the simulations in the NPT ensemble ($P = 6.5\epsilon / {\sigma}^{3}$) with a time step of $dt=0.001\tau$ for $4 \times 10^{7}$ timesteps.
For the first $5 \times 10^{6}$ timesteps, we enforced the cubic box constraint.
Then we relaxed the cubic box constraint to only enforcing orthorhombic box symmetry.

In our $50$ self-assembly simulations with $80, 000$ particles, when the system forms large BCC crystal grains, they are always accompanied by $\gamma$-Brass crystal grains, and they never dominate the entire system.
To rule out the possibility of finite size effects, we also ran stability tests with larger system sizes for BCC (the last two columns of Table \ref{tab:zetterling_stability_result}).
We used the same protocol for the larger systems, and in all of these simulations, the entire system maintained the BCC crystal structure, which we confirmed by visual inspection using the visualization software OVITO \cite{stukowski2009visualization} and plotting the bond-orientational order diagram (BOD) and radial distribution function.
We also checked the particles' LSs using OVITO's built-in Polyhedral Template Matching (PTM) method \cite{larsen2016robust}, and for all of the stability tests of BCC, over $96\%$ particles have BCC LS with an Root Mean Square Deviation (RMSD) cutoff of $0.1$.

\begin{table}[h]
\centering
\caption{Stability test results of the Zetterling system at $P = 6.5\epsilon / {\sigma}^{3}$, $k_{B}T = 0.85\epsilon$. 
All of the crystals are mechanically stable in almost all of the replicas (total number of replicas: $n$). 
The number of simulations in which the system transitioned to a different structure is indicated by $n^{*}$.
In the simulations in which structural transition occurred, $\gamma$-Brass transitioned to BCC and $\alpha$-Mn transitioned to $\gamma$-Brass.
The system sizes ($N$) and densities ($\rho\sigma^{3}$) we used to initialize the ideal crystals are also shown.
}
\label{tab:zetterling_stability_result}
\begin{tabular}{cccccccc}
    \hline
    & $\gamma$-Brass & BCC   & FCC & $\beta$-Mn & $\alpha$-Mn  & BCC & BCC \\ 
    \hline
    N & 17836 & 6750  & 6912  & 14580 & 12528 & 16000 & 85750 \\ 
    $\rho\sigma^{3}$ & 1.004 & 1.054 & 1.009 & 0.9900 & 0.9756 & 1.054 & 1.054 \\ 
    $n$ & 6 & 4 & 4 & 4 & 8 & 4 & 4\\ 
    $n^{*}$ & 1 & 0 & 0 & 0 & 1 & 0 & 0 \\ 
    \hline
\end{tabular}
\end{table}

\addcontentsline{toc}{subsubsection}{Truncated Tetrahedron System}
\subsubsection*{Truncated Tetrahedron System}
For the TT system, we ran hard particle Monte Carlo (HPMC) simulations in both the NPT and NVT ensembles to test the stability of the relevant crystal polymorphs ($\gamma$-Brass, BCC, FCC, $\beta$-Mn, $\alpha$-Mn).
To account for the effective rounding of the shapes due to the AWCA potential \cite{ramasubramani2020mean}, we treated the shapes as spheropolyhedra and set the rounding radius of the shape to be the radius of the contact sphere as defined in Ref. \citenum{ramasubramani2020mean}.
In addition to particles' positions, we also need to ensure that the orientations of the particles are commensurate with the crystal structure of interest, and we use the following protocol to generate initial orientations.
First we initialized the system in the ideal crystal structure in the \textit{NVT} ensemble, fixed the positions of the particles, only enabled the rotational moves and ran for about $1 \times 10^{5}$ MC sweeps so that the particles can explore the orientation phase space.
Then we gradually compressed/expanded the system to the target density (for the NVT ensemble) or a density at which the pressure roughly matches the target pressure (for the NPT ensemble) over $2 \times 10^{5}$ MC sweeps with both the trial translational and rotational moves enabled.
Note that during this process the simulations were still in the NVT ensemble.
For the NPT simulations, we enabled the box length moves once the system is at the target pressure.
We then ran the simulation for at least $1 \times 10^{7}$ MC sweeps or until the structure transitions to a different one.
For each density/pressure of one crystal, we ran at least four replicas.

The stability test results of the NVT runs are summarized in Fig.~\ref{fig:stability}.
All the crystal polymorphs except BCC can remain mechanically stable in all the replicas at the density ($\phi = 0.61 - 0.62$) corresponding to the self-assembly simulation pressure.
Interestingly, for the systems that still exhibit $\gamma$-Brass crystal order towards the end of the stability test simulations, we found with PTM that many particles have adopted LSs of the simple crystals (BCC, FCC and HCP).
In other words, BCC and/or FCC/HCP crystal clusters coexist with $\gamma$-Brass crystal grains at the end of these $\gamma$-Brass stability test runs.
This phenomenon is consistent with our self-assembly simulation results, i.e., particles adopt LSs of various types even though the entire system still exhibits $\gamma$-Brass order.
We obtained the same results with the NPT simulations, which we only used for corroborating the self-assembly simulation results, and thus do not include them here to avoid redundancy.

\addcontentsline{toc}{subsection}{Supplementary Note 2 – Gibbs Free Energy Calculation}
\subsection*{Supplementary Note 2 – Gibbs Free Energy Calculation}\label{method:gibbs}
\begin{figure}[htbp]
    \centering
    \includegraphics[width=1\linewidth]{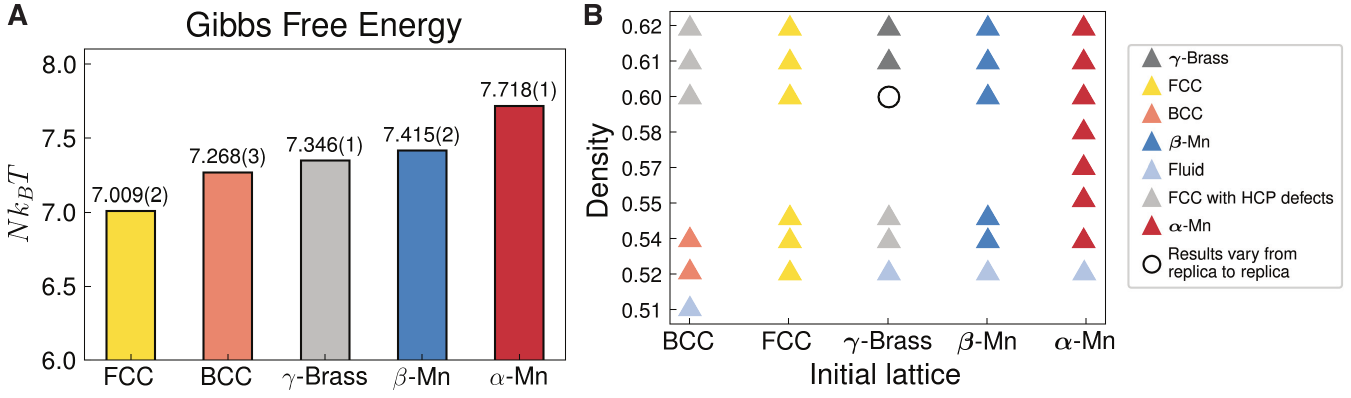}
    \caption{(A) Gibbs free energies of the crystal polymorphs of the Zetterling system.
    (B) Stability test (NVT) results of the TT system.
    The crystal lattices in which the system is initialized are listed on the x-axis.
    For each type of crystal, we used N = 6750 (BCC), N = 6912 (FCC), N=11232 ($\gamma$-Brass), N = 6860 ($\beta$-Mn) and N = 12528 ($\alpha$-Mn) particles.
    The symbols indicate the structure of the entire system at the end of the stability test run.
    For the runs at $\phi = 0.60$ for $\gamma$-Brass, the final structures vary from replica to replica: in some replicas, BCC grains coexist with $\gamma$-Brass grains; in others, the system transitions to FCC.
   }
    \label{fig:stability}
\end{figure}

We computed the Gibbs free energies of all the crystal polymorphs for the Zetterling system at the self-assembly simulation condition ($k_{B}T = 0.85\epsilon$ and $P = 6.5\epsilon / {\sigma}^{3}$).
To compute the Gibbs free energies, we first computed the Helmhotlz free energies using the Einstein molecule method \cite{vega2007revisiting}, which was developed based on the Einstein crystal method proposed by Frenkel and Ladd \cite{frenkel1984new}.
We set the de Broglie wavelength of each particle to unity, the harmonic constant to $1 \times 10^{6} \epsilon/\sigma^{2}$, and ran the simulations in the NVT ensemble with the Langevin thermostat.
The integration was performed in 20 steps, and at each step, the system was equilibrated for $1.5 \times 10^{6}$ timesteps with a time step of $dt = 0.0001\tau$.
For each crystal polymorph, we ran five independent replicas at $k_{B}T = 0.85\epsilon$, density $\rho\sigma^{3}$  with system sizes $N$ ($\rho\sigma^{3}$ and $N$ are summarized in Table~\ref{tab:zetterling_stability_result}).
We estimated the statistical error by computing the standard errors of the five replicas.
Then we gathered the pressure-volume data from the stability tests run in the NPT ensemble, and added the PV data (about $3,000$ data points for each crystal) to the Helmholtz free energies to calculate the Gibbs free energies.
The results are shown in Fig.~\ref{fig:stability}(A).

\addcontentsline{toc}{subsection}{Supplementary Note 3 – Free Energy Profile Calculation}
\subsection*{Supplementary Note 3 – Free Energy Profile Calculation}\label{method:freeE}
To measure the free energy barrier for the TT system, we performed $20$ independent biased metadynamics simulations with $N = 2000$ particles using the hybridMC-MetaD algorithm ~\cite{zhao2026hybrid}.
We constructed the TT shape following the same protocol detailed in the \hyperref[method:sim]{Simulation Protocols} section in the main text, and used the same parameters for the AWCA potential.

For the hybridMC-MetaD simulations, we used the Steinhardt order parameter $Q_{6}$ as the biased collective variable.
To determine the neighbor list for $Q_{6}$, we used a cutoff radius of $6.8\sigma$.
We initialized the system by placing the particles in a cubic array at a low density ($\rho\sigma^{3} = 0.33$) in a cubic simulation box, and thermalized the system at $k_{B}T = 1\epsilon$ for $1 \times 10^{5}$ timesteps in the canonical ensemble.
Then the system was compressed to the target density, $\rho\sigma^{3} = 0.615$, over $1 \times 10^{5}$ timesteps, which corresponds to the pressure at which we ran the self-assembly simulations.
Using the resulting configuration at the target density, we ran the hybridMC-MetaD simulations for $2 \times 10^{8}$ timesteps.
The hybridMC-MetaD parameters are listed in Table \ref{tab:metaD_params}, and the types of transitions observed and their frequencies are listed in Table \ref{tab:metaD_results}.

\begin{figure}[htbp]
    \centering
    \includegraphics[width=0.4\linewidth]{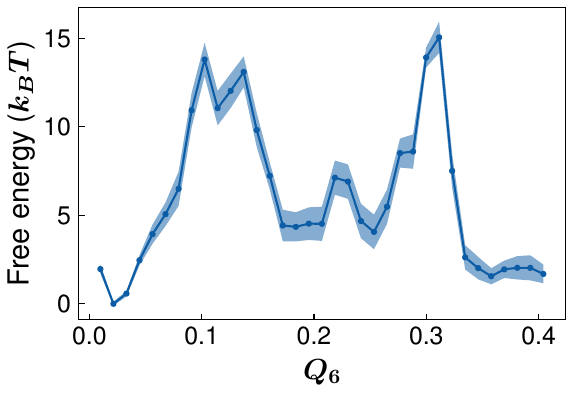}
    \caption{Free energy profile along $Q_{6}$ for the TT system, constructed by averaging over $20$ independent hybridMC-MetaD simulations. Shading of the curve indicates the width of error bars. The first well at $Q_6 < 0.05$ corresponds to the fluid phase. The second well at $Q_6 \approx 0.18$ corresponds to $\gamma$-Brass. The third well at around $Q_6 \approx 0.25$ corresponds to mixture phases that contain LSs of $\gamma$-Brass, BCC, and FCC LSs. The fourth well at $Q_6 > 0.3$ corresponds to FCC with HCP defects.
   }
    \label{fig:1D_freeE}
\end{figure}

To construct the free energy profile, we reweighted the data from each simulation and averaged over the $20$ independent runs.
The free energy profile along $Q_{6}$ is shown in Fig.~\ref{fig:1D_freeE}.
The first well at $Q_6 < 0.05$ corresponds to the fluid phase. The second well at $Q_6 \approx 0.18$ corresponds to $\gamma$-Brass. The third well at around $Q_6 \approx 0.25$ corresponds to mixture phases that contain LSs of $\gamma$-Brass and BCC, and in some simulations, small amount of FCC LSs are also present. Finally, the energy well at $Q_6 > 0.3$ corresponds to FCC with HCP defects. Therefore, the system needs to overcome a free energy barrier of around $12 k_BT$ to form $\gamma$-Brass from the fluid phase, while the energy barrier for the system to form FCC from $\gamma$-Brass is only around $10 k_BT$, and the energy barrier for $\gamma$-Brass to develop LSs of BCC or FCC is almost negligible.

\begin{table}[h]
\centering
\caption{hybridMC-MetaD simulation parameters for TT system. CV is the collective variable. $L$ is the number of timesteps in the short MD run in the hybridMC-MetaD simulation. $\tau_{\text{MetaD}}$ is the bias potential update interval. $\tau_{\text{BDP}}$ is the time constant for the Bussi-Donadio-Parrinello thermostat. $\omega_{0}$ and $\sigma_{G}$ are the initial height and width of the Gaussian bias potentials. $\gamma$ is the bias factor of well-tempered metadynamics.
}
\label{tab:metaD_params}
\begin{tabular}{lllcccll}
\toprule
CV   &$dt$& $L$& $\tau_{\text{MetaD}}$& $\tau_{\text{BDP}}$ & $\omega_{0}$& $\sigma_{G}$ & $\gamma$ \\
\midrule
$Q_{6}$  &0.005 &100 &500$Ldt$ &100$dt$ &0.5 &0.002 &20 \\
\bottomrule
\end{tabular}
\end{table}

\begin{table}[h]
\centering
\caption{Types of transitions observed in the $20$ hybridMC-MetaD simulations and their frequencies.
}
\label{tab:metaD_results}
\begin{tabular}{lc}
\toprule

Transition   & Number of simulations \\
\midrule
fluid $\rightarrow$ $\gamma$-Brass  & 2 \\
fluid $\rightarrow$ $\gamma$-Brass $+$ BCC  & 1 \\
fluid $\rightarrow$ $\gamma$-Brass $\rightarrow$ FCC  & 3 \\
fluid $\rightarrow$ $\gamma$-Brass $+$ BCC $\rightarrow$ FCC  & 3 \\
fluid $\rightarrow$ FCC  & 7 \\
fluid $\rightarrow$ BCC $\rightarrow$ FCC & 2 \\
fluid $\rightarrow$ BCC $+$ FCC & 2 \\

\bottomrule
\end{tabular}
\end{table}

\addcontentsline{toc}{subsection}{Supplementary Note 4 – Local Structure Classification}
\subsection*{Supplementary Note 4 – Local Structure Classification}

\addcontentsline{toc}{subsubsection}{Training Dataset Selection}
\subsubsection*{Training Dataset Selection} \label{method:training_data}
For both systems, we used the same set of features, and we list the parameters we used in Table \ref{tab:features}.

\begin{table}[h]
\centering
\caption{Parameters for computing the local structure features with \textit{freud}.
Note that we do not list the parameters for Voronoi density here as no parameters are needed.}
\label{tab:features}
\begin{tabular}{cccccc}
\hline
Order parameter& $l$ values& \texttt{weighted}&\texttt{average}&\texttt{wl}&\texttt{wl\_normalize}\\ 
\hline
$q'_{l}$ & 4, 6, 7, 8, 9, 10, 12& \texttt{True}& \texttt{False}& \texttt{False}&\texttt{False}\\
 $w_{l}$ & 4, 6, 8, 10, 12& \texttt{False}& \texttt{False}& \texttt{True}&\texttt{True}\\
 $\bar{q'_{l}}$ & 4, 6, 7, 8, 9, 10, 12& \texttt{True}& \texttt{True}& \texttt{False}&\texttt{False}\\ 
$\bar{w_{l}}$ & 4, 6, 8, 10, 12& \texttt{False}& \texttt{True}& \texttt{True}&\texttt{True}\\
\hline
\end{tabular}
\end{table}

\paragraph{Zetterling System}
For the Zetterling system, we used data from the stability test runs in the NPT ensemble at the self-assembly simulation condition.
Although we ran multiple tests with different random seeds for each crystal polymorph, the simulation trajectories of the different replicas are almost identical both in terms of global structures and the LSs, except the ones in which structural transitions occurred (Table \ref{tab:zetterling_stability_result}).
We checked the global structures using methods detailed in the \hyperref[method:global]{Crystal Structure Identification} section in main text for selected frames along the simulation trajectory for each replica.
For these frames, we also computed the distributions of the MSMs to check whether the LSs vary along the trajectories.
We found that the distributions remained almost the same throughout the trajectories, and when we compared the distributions across the replicas, we found no significant difference either.
Therefore, the particular choice of replicas or frames should not affect the performance of our classifiers.

Due to the nature of $k$NN, however, the number of samples in each class in the training dataset, i.e., the balance of the classes, can have a significant impact on $k$NN classifiers' ability to accurately identify the instances \cite{beckmann2015knn}.
In our preliminary study of the two systems, we found that the balance of the complex crystal classes and the fluid class greatly affect the amount of particles classified as having these types of LSs \textit{before} crystallization.
For example, if we have more fluid data in the training dataset, at the early stages of the pathways, more particles will be labeled as fluid-like instead of complex-crystal-like.
We note that this is due to the innate similarity of the LSs of the complex crystals and fluid.
In addition, before the system forms any crystalline order, the complex-crystal LSs deviate even more from crystalline-like and thus it is harder to distinguish them from the fluid LSs.
As we do not know the ground truth of the exact types of particles' LSs along the crystallization pathways, and we see similar good accuracy scores when we evaluated the classifiers on the stability test data (see more details in the \hyperref[method: validation]{Validation} section),  we adjusted the number of samples in each class based on our knowledge of the system, and utilized $k$NN's sensitivity to dataset balance to our advantage by incorporating samples for each class based on the frequency of the crystals (Fig.~\ref{fig:polymorphism}(B) in main text).
For example, we included more samples for the fluid class for both systems (Table \ref{tab:training_dataset}) to account for the constant presence of defects in the crystal, which we found when inspecting the simulation trajectories directly.
As we have not observed $\alpha$-Mn for the Zetterling system (or $\beta$-Mn for the TT system), we included less $\alpha$-Mn (or $\beta$-Mn) data in the training dataset for the Zetterling system (or TT system).

The number of samples in the simple crystal LS classes has a much less significant effect on the classification results because these LSs are easier to distinguish from those in the complex crystals and fluid, and almost no particles have simple-crystal LSs when the system is still disordered.
In addition, we used OVITO's built-in PTM method to identify the simple-crystal LSs and confirmed our classification results (see the \hyperref[method: validation]{Validation} section for more details).
We note that the balance of the classes does not affect the classification results \textit{after} crystallization as much, especially for the Zetterling system, as the entire system tends to transition to one dominant crystal quickly, and indeed almost all of the particles in the crystals are classified as having the corresponding LSs.

The Zetterling system can form large crystal grains of BCC, FCC, $\gamma$-Brass and $\beta$-Mn, so we selected 10 frames from one stability test simulation as the training data for each of these structures.
We have never observed $\alpha$-Mn in any of our self-assembly simulations, so we only used 2 frames for $\alpha$-Mn.
For HCP, we took the last 5 frames of several self-assembly trajectories in which the system formed FCC (with HCP defects) and used PTM to identify the HCP-like particles.
Then we used them as the training data for the HCP class.
As the HCP crystal grains form less frequently than the other crystals, and HCP-like particles usually only appear as defects, we used a smaller dataset for HCP.
For the fluid class, to include more configurations, we used ten first frames of the self-assembly simulation trajectories of the $8,000$ particle system as the training data\footnote[1]{If we used frames from the simulation trajectories of the $80,000$ particle system, we would only have been able to use two frames/configurations.}.
The number of samples we used for each class in the training datasets is summarized in Table \ref{tab:training_dataset}.

For the preprocessing step, we scaled the features to the range $(-5, 5)$ for the fluid/crystal classifier and performed principal component analysis (PCA) on the features for the crystal LS classifier.
\begin{table}[h]
\centering
\caption{The number of samples in each class in the training datasets for the Zetterling system and the TT system.}
\label{tab:training_dataset}
\begin{tabular}{ccc}
\hline
Class& Zetterling system& TT system\\
\hline
 $\gamma$-Brass& 65000& 51151\\
 BCC& 67500& 55543\\
 FCC& 69120& 55296\\
 HCP& 57439& 30245\\
 $\beta$-Mn& 68600& 27440\\
 $\alpha$-Mn& 25056& 50112\\ 

Fluid& 80000& 70000\\
\hline
\end{tabular}
\end{table}

\paragraph{Truncated Tetrahedron System}
As mentioned in the \hyperref[method:stability]{Crystal Stability Test} section, we always initialize the stability tests of the TT system in the NVT ensemble and the particles tend to adopt LSs of other crystal structures as we run the simulations longer.
Therefore, we only used data from the initial stage of the stability test (immediately after we switched on the trial translational and rotational moves), i.e., the NVT part of the simulation as the training data.

As explained above, the balance of the classes affects the classification results, and we select the number of samples (summarized in Table \ref{tab:training_dataset}) following the same rules we used for the Zetterling system.
Given that BCC is not mechanically stable at densities close to our self-assembly simulation condition ($P=$ 0.32 or $\phi= 0.61 - 0.62$) but lower densities (Fig.~\ref{fig:stability}), we used 4 replicas run at $\phi = 0.54$, and identified the BCC-like particles using PTM (RMSD cutoff = 0.15).
Then from each simulation trajectory, we selected 5 frames with the highest number of BCC-like particles, and computed the MSMs for these particles as the BCC training data.
For FCC, $\beta$-Mn, $\alpha$-Mn and $\gamma$-Brass, we used the first several frames of different replicas run at $\phi = 0.62$ as our training data.
In addition, for $\gamma$-Brass, as mentioned above, about $20\% - 25\%$ particles of the system tend to form simple crystal LSs (BCC, FCC and HCP) as identified by PTM (RMSD cutoff = 0.15), so we removed these particles from the $\gamma$-Brass simulation frames.
For HCP, we used PTM (RMSD cutoff = 0.11) on all the self-assembly simulation trajectories in which the system assembled into FCC to identify the HCP-like particles, and then we used the frames with the highest number of HCP-like particles as the HCP training data.
For the fluid phase, we selected $7$ self-assembly trajectories in which the system assembled into different structures and used the first frames as the training data.

For the preprocessing step, we standardized the features by removing the mean (we do not scale the variance of the data) for the fluid/crystal classifier and performed principal component analysis (PCA) on the features for the crystal LS classifier.

\addcontentsline{toc}{subsubsection}{Validation}
\subsubsection*{Validation} \label{method: validation}
We used multiple methods to evaluate our classifiers.
First we estimated the accuracy of the classifiers with different $k$ values by performing  $5$-fold cross-validation on our training datasets.
The mean scores and standard deviations are reported in Table \ref{tab:zetterling_score} and Table \ref{tab:tt_score}.
As can be seen, the classifiers show consistent good performance on the training dataset, regardless of the $k$ values.
We note that the high accuracy score is because we used thermalized ideal crystals as the training data for the crystal LSs, and the local structure features we selected are sufficient for quantifying the LSs and distinguishing the complex-crystal LSs from the fluid ones.
Because of this, the balance of the classes in the training dataset does not affect the performance of the classifiers as much when they are evaluated on the thermalized ideal crystal data.

Compared with the thermalized ideal crystals, the self-assembled crystals have more defects and thermal noise, and thus the accuracy scores estimated on the training datasets may not be reflective of the true performance of the classifiers when they are used for classifying the LSs along the crystallization pathways.
However, we can not use data from the crystallization pathways for the crystal LSs in the training dataset due to the lack of ground truth for the LS types.
Therefore, we also evaluated the classification results based on our knowledge and physical understanding of the two systems.
For example, we colored the particles based on their assigned classes and inspected the results with OVITO, and compared the results for the different $k$ values listed in Tables \ref{tab:zetterling_score} \& \ref{tab:tt_score}.
This way, we could estimate the rough range for $k$.
Similar to the accuracy scores, we did not see significant difference in the results, and any value of $k$ close to the ones we chose in Tables \ref{tab:zetterling_score} and \ref{tab:tt_score} should produce similar classification results and thus same conclusions.

Furthermore, we identified the simple-crystal LSs along the crystallization pathways using the PTM method and compared with the results of our classifiers and found a good qualitative match between the two.

\begin{table}[h]
\centering
\caption{Mean accuracy scores of the classifiers for different $k$ values (Zetterling system).
For the fluid/crystal classifier, we used $k = 11$, and for the crystal LS classifier we used $k = 55$.}
\label{tab:zetterling_score}
\begin{tabular}{ccc}
\hline
$k$& Fluid/crystal classifier& Crystal LS classifier\\
\hline
 \textbf{11}& \textbf{0.961(1)}& 0.989(0)\\
 25& 0.960(1)& 0.988(0)\\
 \textbf{55}& 0.957(0)&\textbf{0.987(0)}\\
 65& 0.957(0)& 0.987(0)\\ 

75& 0.956(1)& 0.986(0)\\
\hline
\end{tabular}
\end{table}

\begin{table}[h]
\centering
\caption{Mean accuracy scores of the classifiers for different $k$ values (TT system).
For the fluid/crystal classifier, we used $k = 15$, and for the crystal LS classifier we used $k = 55$.}
\label{tab:tt_score}
\begin{tabular}{ccc}
\hline
$k$& Fluid/crystal classifier& Crystal LS classifier\\
\hline
 \textbf{15}& \textbf{0.935(1)}& 0.956(1)\\
 25& 0.936(1)& 0.955(1)\\
 \textbf{55}& 0.935(1)&\textbf{0.952(1)}\\
 65& 0.935(1)& 0.951(1)\\ 

75& 0.934(1)& 0.950(1)\\
\hline
\end{tabular}
\end{table}

\clearpage

\addcontentsline{toc}{section}{Supplementary Figures}
\section*{Supplementary Figures}

\begin{figure}[h]
    \centering
    \includegraphics[width=1\linewidth]{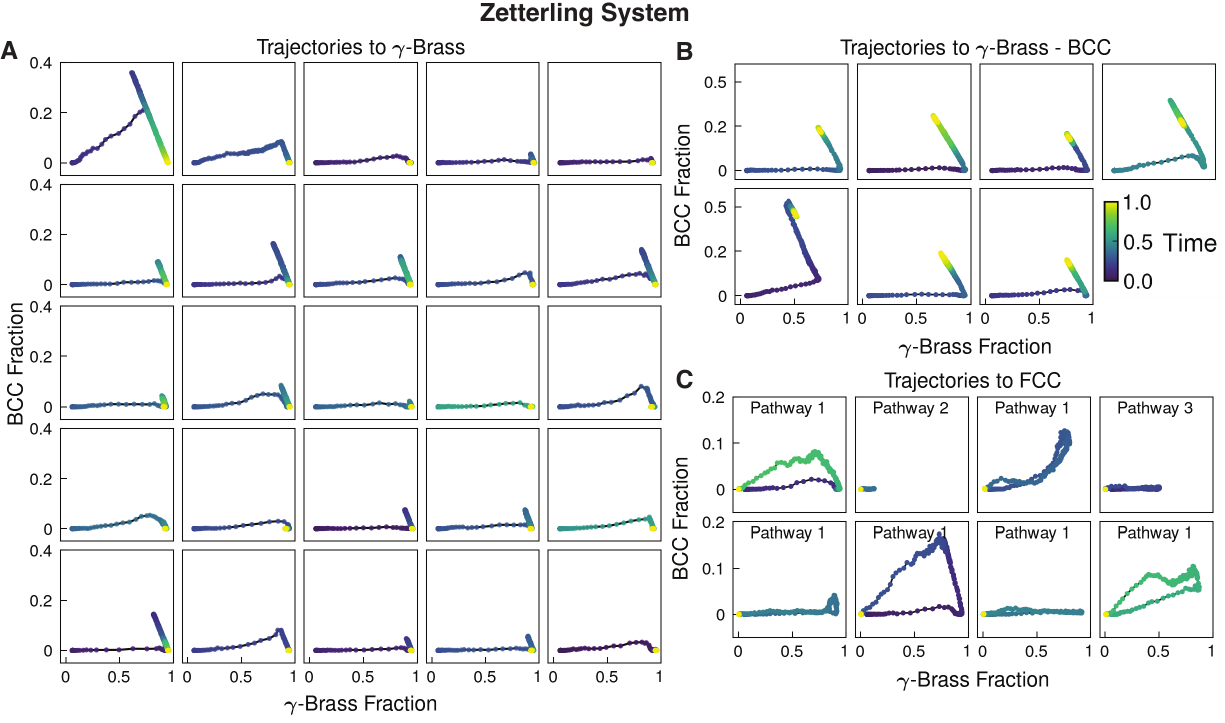}
    \caption{(Zetterling system) Fraction of BCC-like particles plotted against fraction of $\gamma$-Brass-like particles for trajectories in which the system formed:
   (A) $\gamma$-Brass,
   (B) mixture of $\gamma$-Brass and BCC, and
   (C) FCC.
   For each trajectory, the time duration is scaled to a range from 0 to 1, and the color of the markers indicates timestep.
}
    \label{fig:zetterling_bcc_brass}
\end{figure}

\begin{figure}[h]
    \centering
    \includegraphics[width=1\linewidth]{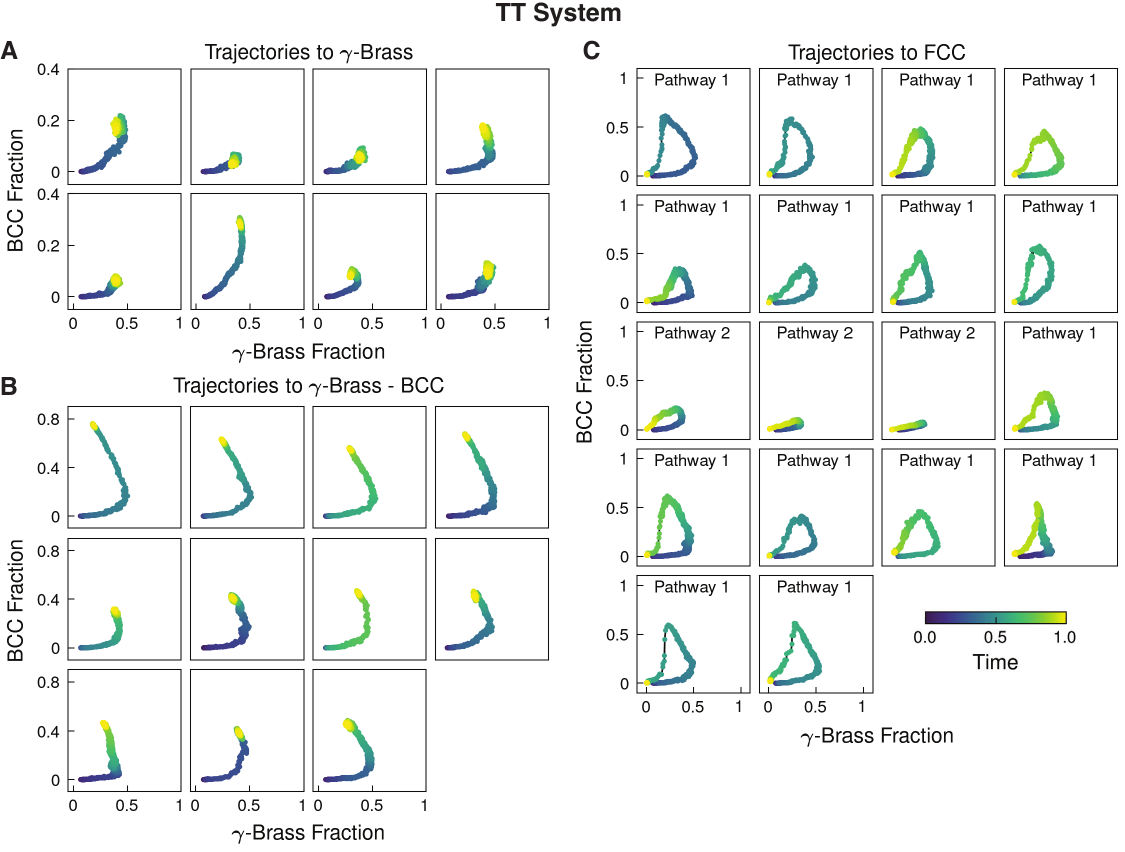}
    \caption{(TT system) Fraction of BCC-like particles plotted against fraction of $\gamma$-Brass-like particles for trajectories in which the system assembled into:
   (A) $\gamma$-Brass,
   (B) mixture of $\gamma$-Brass and BCC, and
   (C) FCC.
   For each trajectory, the time duration is scaled to a range from 0 to 1, and the color of the markers indicates timestep.
}
    \label{fig:tt_bcc_brass}
\end{figure}

\begin{figure}[h]
    \centering
    \includegraphics[width=1\linewidth]{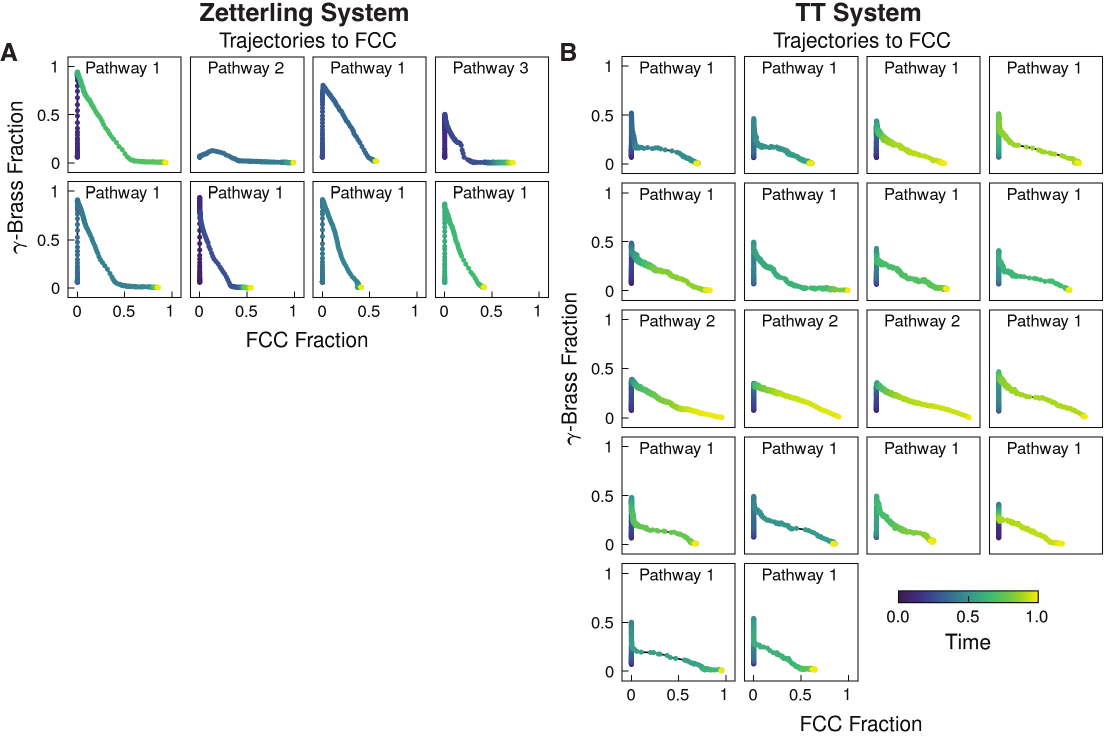}
    \caption{Fraction of $\gamma$-Brass-like particles plotted against fraction of FCC-like particles for trajectories in which the (A) Zetterling system and (B) TT system formed FCC.
   For each trajectory, the time duration is scaled to a range from 0 to 1, and the color of the markers indicates timestep.}
    \label{fig:fcc_rbrass}
\end{figure}

\begin{figure}[h]
    \centering
    \includegraphics[width=1\linewidth]{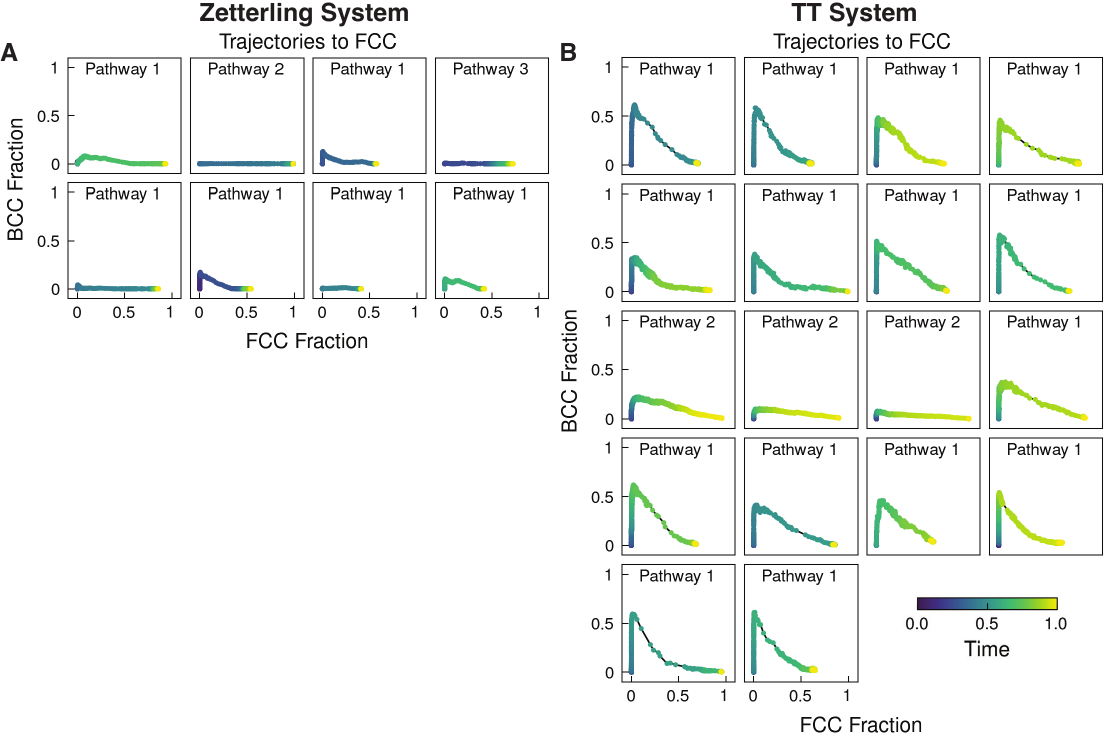}
    \caption{Fraction of BCC-like particles plotted against fraction of FCC-like particles for trajectories in which the (A) Zetterling system and (B) TT system formed FCC.
   For each trajectory, the time duration is scaled to a range from 0 to 1, and the color of the markers indicates timestep.}
    \label{fig:fcc_bcc}
\end{figure}

\begin{figure}[h]
    \centering
    \includegraphics[width=1\linewidth]{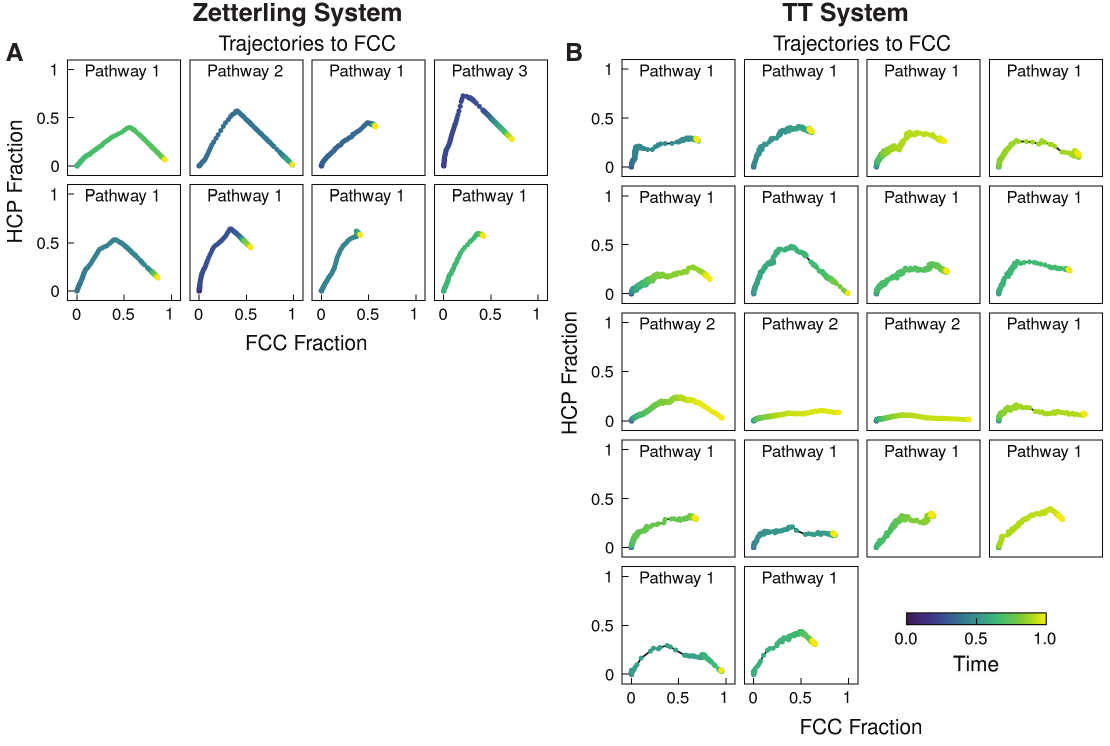}
    \caption{Fraction of HCP-like particles plotted against fraction of FCC-like particles for trajectories in which the (A) Zetterling system and (B) TT system formed FCC.
   For each trajectory, the time duration is scaled to a range from 0 to 1, and the color of the markers indicates timestep.}
    \label{fig:fcc_hcp}
\end{figure}

\begin{figure}[h]
    \centering
    \includegraphics[width=0.8\linewidth]{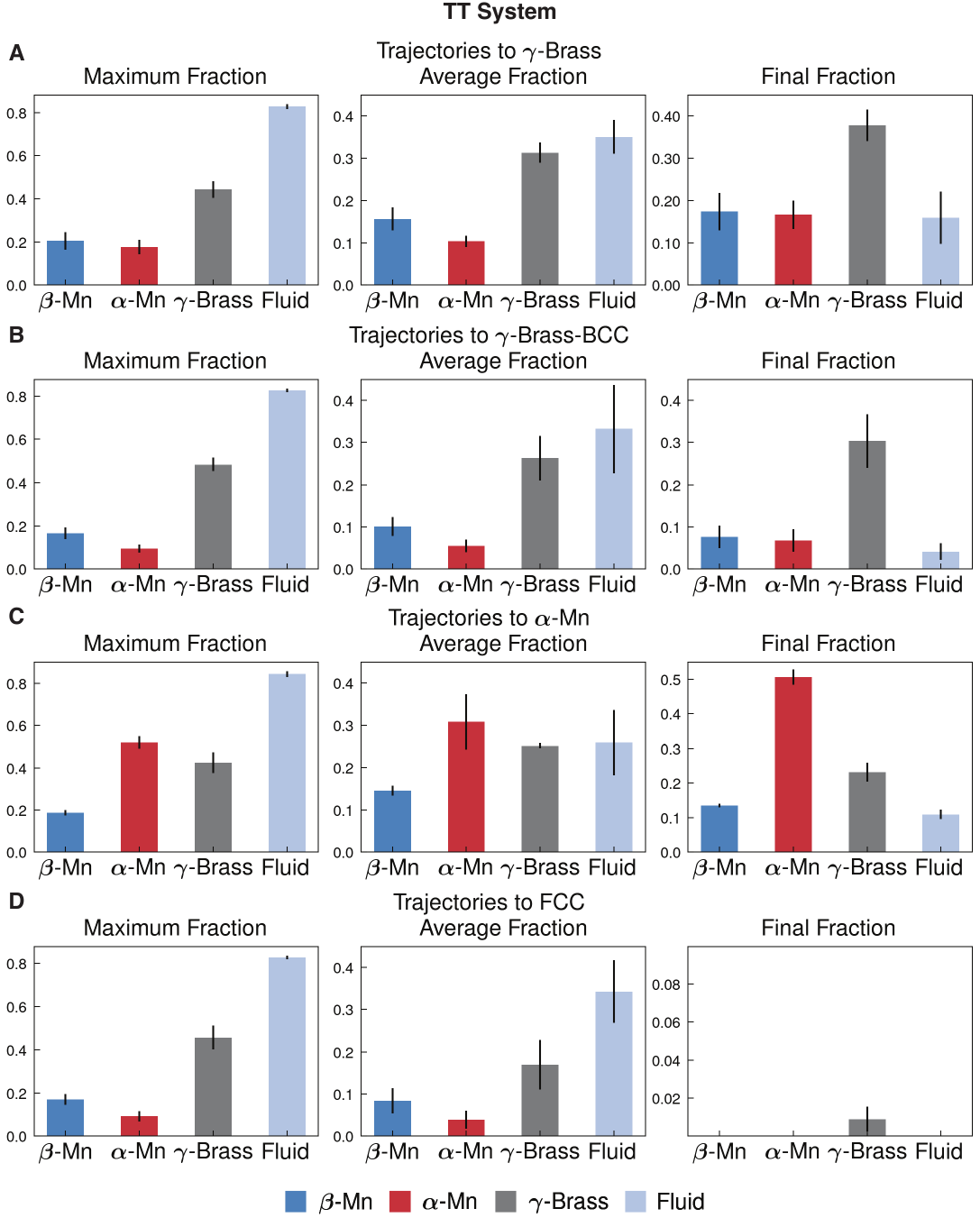}
    \caption{(TT system) Fractions of particles that have complex-crystal LSs ($\beta$-Mn, $\alpha$-Mn, $\gamma$-Brass) and fluid LSs.
    In each row, from left to right:
    (left) maximum fractions of the LSs of the entire simulation trajectory,
    (center) average fractions of the LSs of the entire trajectory, and
    (right) fractions of LSs in the system at the end of the simulation trajectory (averaged over the last 10 frames of the trajectory).
    The height of the bar represents the averaged value calculated using all the trajectories in which the system assembled into the corresponding structure.
    We used (A) $8$, (B) $11$, (C) $2$ and (D) $18$ trajectories for each structure.
    The standard deviation is drawn as a black line at the top of the bar.
}
    \label{fig:complex-env-shape}
\end{figure}

\begin{figure}[h]
    \centering
    \includegraphics[width=0.8\linewidth]{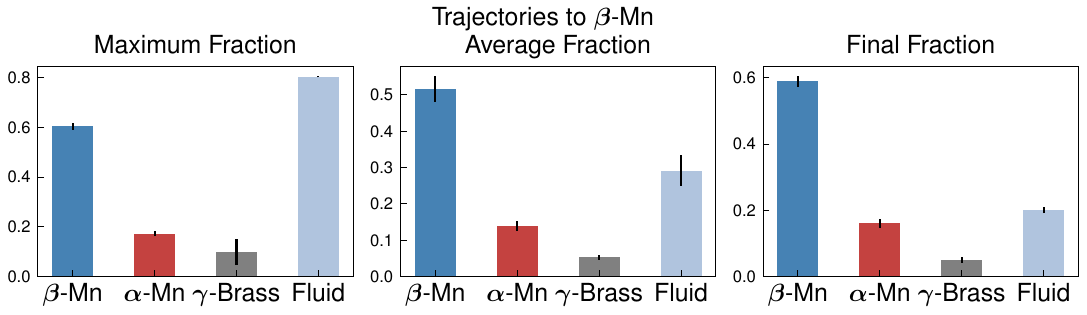}
    \caption{(Zetterling system) Fractions of particles that have complex-crystal LSs ($\beta$-Mn, $\alpha$-Mn, $\gamma$-Brass) and fluid LSs in the $\beta$-Mn trajectories.
    From left to right:
    (left) maximum fractions of the LSs of the entire simulation trajectory,
    (center) average fractions of the LSs of the entire trajectory, and
    (right) fractions of LSs in the system at the end of the simulation trajectory (averaged over the last 10 frames of the trajectory).
    The height of the bar represents an averaged value calculated using all $10$ trajectories in which the system formed $\beta$-Mn. The standard deviation is drawn as a black line at the top of the bar.
}
    \label{fig:complex-env-mn-zetterling}
\end{figure}

\clearpage

\addcontentsline{toc}{section}{References}
\bibliography{sn-bibliography}

\makeatletter
\@input{main_aux.tex}
\makeatother